\documentclass[fleqn,usenatbib]{mnras}

\usepackage{newtxtext,newtxmath}

\makeatletter
\DeclareFontEncoding{LS1}{}{}
\DeclareFontSubstitution{LS1}{stix2}{m}{n}
\DeclareSymbolFont{stix2letters}{LS1}{stix2}{m}{it}
\SetSymbolFont{stix2letters}{bold}{LS1}{stix2}{b}{it}
\AtBeginDocument{
  \DeclareMathSymbol{v}{\mathalpha}{stix2letters}{`v}
}
\makeatother

\usepackage[T1]{fontenc}

\DeclareRobustCommand{\VAN}[3]{#2}
\let\VANthebibliography\thebibliography
\def\thebibliography{\DeclareRobustCommand{\VAN}[3]{##3}\VANthebibliography}

\usepackage{graphicx}	
\usepackage{amsmath}	

\title[Protoplanetary disc evolution with magnetic winds]{Secular evolution of viscous and self-gravitating protoplanetary discs with magnetic  winds}

\author[E. R. Redkin et al.]{
Evgenii R. Redkin,$^{1, 2}$
Eduard I. Vorobyov,$^{3,1}$
Paola Pinilla,$^{4}$\thanks{E-mail: p.pinilla@ucl.ac.uk}
Fabiola A. Gerosa,$^{4}$ 
and Konstanze Zwintz$^3$
\\
$^{1}$Research Institute of Physics, Southern Federal University, Stachki Ave. 194, 344090 Rostov-on-Don, Russia\\
$^{2}$Faculty of Physics, Southern Federal University, Zorge 5, 344090 Rostov-on-Don, Russia\\
$^{3}$Institut für Astro- und Teilchenphysik, Universität Innsbruck, Technikerstraße 25, 6020 Innsbruck, Austria\\
$^{4}$Mullard Space Science Laboratory, University College London, Holmbury St Mary, Dorking, Surrey RH5 6NT, UK
}

\date{Accepted 2026 September 15. Received 2026 September 5; in original form 2026 June 17}

\pubyear{\the\year{}}

\begin{document}
\label{firstpage}
\pagerange{\pageref{firstpage}--\pageref{lastpage}}
\maketitle

\begin{abstract}

Traditionally, the disc is believed to evolve under the influence of turbulent viscosity, although the importance of this mechanism has recently been questioned. Alternatively, magnetic disc wind or gravitational instability could be considered, but which of these three mechanisms contributes the most at a given evolutionary stage remains an open question. We present global numerical hydrodynamic simulations of protoplanetary disc evolution in the thin-disc limit on disc lifetime timescales, including all three mechanisms of mass and angular momentum transport. We calculate the gravitational, viscous, and magnetic torques to assess the contribution of each mechanism. We found that these transport mechanisms occupy distinct zones of influence in the disc depending on the stage of evolution: the dominance of disc self-gravity during the early phase is replaced by the prevalence of magnetic disc wind and viscosity, with the latter being especially relevant at the disc periphery. We show that a noticeable decrease in disc mass occurs only after the termination of the embedded phase. The suppression of disc viscous spreading is achieved only for the most intense wind; otherwise, the disc continues to grow in size till the end of simulations. The spiral structure contributes to the temporary retention of dust in the disc. However, this is insufficient to prevent the depletion of dust on long timescales. The results emphasize the importance of a comprehensive approach to long-term simulations of protoplanetary discs with no single mechanism of mass and angular momentum transport regarded as exclusive.

\end{abstract}

\begin{keywords}
protoplanetary discs -- accretion, accretion discs 
\end{keywords}



\section{INTRODUCTION}
\label{Section: introduction}

The formation of a protostar results from the gravitational collapse of a cold, slowly rotating molecular cloud. The mass of the young stellar object continues to increase through the accretion of matter from the protoplanetary disc, which forms around the protostar due to the conservation of angular momentum of the parent cloud \citep{Larson2003}. Over time, under the influence of various physical processes, the spatial structure, size, and mass of the protoplanetary disc change, which in turn can affect the architecture of planetary systems. Therefore, understanding exactly how protoplanetary discs evolve under the action of various mechanisms of transport and loss of mass and angular momentum is the key to constrain the theories of planet formation \citep{Morbidelli2016}.

During the initial stages of the embedded phase, strong influence on the evolution of a young, massive protoplanetary disc is exerted by gravitational instability \citep{Vorobyov2009,Kratter2016}. The continuing influx of matter from the collapsing cloud during this period replenishes the disc mass lost through accretion onto the star, maintaining the disc in a gravitationally unstable state \citep{Kratter2008,Rice2010,Longarini2025} and promoting the development of disc fragmentation \citep{VorobyovBasu2005}. Gravitational torques are predominantly negative and drive the matter inward and angular momentum outward \citep{Vorobyov2007,Das2026}.

The question of which mechanism determines disc evolution at later stages remains open. In the seminal works of \citet{LyndenBell&Pringle1974, Pringle1981}, the idea was formulated that turbulence in protoplanetary discs can act as an effective viscosity that transports angular momentum outward, while the bulk of the mass moves inward and accretes onto the star. This classical approach, the so-called ``viscous evolution'', is widely used in the study of protoplanetary discs \citep{Nakamoto&Nakagawa1994, Hartmann1998, Birnstiel2010, Trapman2020}. However, doubts have recently arisen as to whether the level of turbulence in protoplanetary discs is high enough to reproduce the observed accretion rates and the characteristic disc lifetime of several million years \citep{Rosotti2023}. Magneto-rotational instability \citep[MRI,][]{Balbus1991}, which is usually considered the main source of turbulence, may not operate in most of the disc due to a low degree of ionization, except from the innermost regions and upper layers \citep{Gammie1996}. Moreover, when non-ideal magnetohydrodynamic (MHD) effects, such as ohmic resistivity and ambipolar diffusion, are taken into account, simulations show significantly weaker overall turbulence (see \citet{Lesur2021} and references therein). Evidence against strong turbulence is provided not only by theoretical works \citep{Turner2014, Mori2016}, but also by observations, in particular, by the velocity dispersion derived from CO lines \citep{Flaherty2015, Flaherty2020}, as well as by observations of the width and thickness of dust rings in protoplanetary discs \citep{Pinte2016, Franceschi2023}. However, recent observations do not rule out strong turbulence in discs of IM Lup, HL Tau and several others \citep{Paneque-Carreno2024, Sai2026, Antilen2026}, so the level of turbulence is still a subject of debate.

As an alternative mechanism driving the disc evolution under conditions of extremely weak turbulence, a magnetized disc wind can be considered, which arises if the magnetic field lines are sufficiently inclined to the disc normal \citep{Blandford1982}. Matter can start to move along such a line, accelerating under the action of centrifugal force and, accordingly, increasing its angular momentum. Thus, the magnetized disc wind leads to the removal of mass and angular momentum from the system, and also exerts a torque that acts on the matter in the disc and promotes its inward motion. In doing so, the magnetized disc wind not only stimulates the transport of matter through the disc and protostellar accretion by carrying away angular momentum (wind-driven accretion), but also draws energy from this accretion flow, since the magneto-centrifugal mechanism for launching the wind is a process of converting the gravitational energy of the disc matter into the kinetic energy of the outflowing material \citep{Frank2004}.

Intensive development of this approach began with the advent of numerical 3D MHD simulations, first using local shearing boxes \citep{Bai2013, Simon2013, Lesur2014}, and later global models that investigated the magnetized disc wind at high numerical resolution \citep{Gressel2015, Bethune2017, Bai2017, Wang2019}. However, direct 3D modeling of disc evolution over timescales on the order of several million years remains computationally inaccessible owing to high computational cost. In this regard, one-dimensional (semi-)analytical models have become widespread, making it possible to investigate long-term disc evolution under the influence of the wind  \citep{Armitage2013, Bai2016, Suzuki2016, Chambers2019, Kunitomo2020, Lesur2021-1D, Alessi&Pudritz2022, Tabone2022, Weder2023}. Despite the usefulness of such one-dimensional axisymmetric models, they have limitations. In particular, they do not allow for the self-consistent treatment of gravitational instability at the early stages of disc evolution, and rarely provide the possibility of direct comparison of the contributions of viscous and magnetic wind transport mechanisms within a single computational framework.

For purely viscous evolution and evolution driven solely by a magnetized wind, the dependence of both gas and dust disc sizes on time may differ significantly: viscous discs expand over time, whereas the radii of discs evolving under the action of the wind decrease \citep{Trapman2022, Zagaria2022}. \citet{Manara2023}  noted that these differences may help differentiating the disc evolution mechanisms on the basis of observational data.
However, in real protoplanetary discs, both mechanisms probably act simultaneously or at different evolutionary stages, and self-gravity also plays an important role in young discs.

In the context of the development of simplified models of magnetic wind-driven accretion, \citet{Kadam2025} presented a model of protoplanetary disc evolution with a self-consistent description of the wind through approximation formulas obtained from local shearing box simulations of \citet{Bai2013-2}. In the present work, an alternative approach is used, based on the $\alpha$-parameterization of \citet{Tabone2022}, which provides an effective description of the wind's impact on the disc, resembling the classical $\alpha$-parameter \citet{ShakuraSunyaev1973} in the viscous theory of accretion discs. The use of a simpler magnetized disc wind model with a constant value of $\alpha_{\rm wind}$ enables a direct comparison with the contribution of turbulent viscosity to the global disc evolution.

In this work, based on numerical hydrodynamic simulations in the thin-disc limit using the Formation and Evolution Of Stars And Discs (FEOSAD) code, the long-term disc evolution under the combined action of all three angular momentum transport mechanisms -- gravitational instability, viscosity, and magnetized wind -- is investigated. Unlike one-dimensional models, our two-dimensional ($r,\phi$) numerical model allows for the self-consistent calculation of gravitational torques and a direct comparison of the contribution of each mechanism considered.
We particularly focus on the influence of the magnetic disc wind on the size and mass of discs over timescales of up to 2 Myr, comparable to the characteristic lifetimes of protoplanetary discs, in the presence of non-negligible turbulent viscosity and disc self-gravity.

This article is structured as follows. Section~\ref{Section: model_description} describes the numerical model used in the work. Section~\ref{Section: results} presents the main results of the simulations: the evolution of disc masses and sizes is analysed, a comparison of gravitational, viscous, and magnetic torques is carried out, and the problem of catastrophic dust disc depletion is considered separately.
In Section~\ref{Section: discussion}, the limitations of the model are outlined and a comparison with other published works is made.  
Section~\ref{Section: conclusions} presents the main conclusions of the work.

\section{MODEL DESCRIPTION}
\label{Section: model_description}

This work is based on numerical hydrodynamic simulations performed using the FEOSAD code. 
A detailed description of this numerical model is presented in \citet{Vorobyov2018}, as well as in \citet{Vorobyov2020_Effect, Vorobyov2023}, where the back-reaction of dust on gas and the modification of the gas-dust drag force taking into account the nonlinear Stokes regime were added.

The simulations were carried out on a two-dimensional ($r, \phi$) polar grid containing $256 \times 256$ cells, with logarithmic spacing in the radial direction and linear spacing in the azimuthal direction. The hydrodynamic equations are solved in the thin-disc limit for gas, grown dust, and small dust. The integration of the hydrodynamic equations was performed using a finite-volume method with a time-explicit solution procedure, methodologically similar to the ZEUS code \citep{StoneNorman1992}. The advection of gas and dust was carried out using a third-order accurate piecewise-parabolic interpolation scheme \citep{Colella1984}.
Cooling and heating were computed using an implicit Newton-Raphson integrator. The friction between gas and dust, including back-reaction, was calculated using a semi-analytic solution scheme \citep{LorenAguilar2015, Stoyanovskaya2018}.

\subsection{Gas component}
\label{Subsection: gas_component}

The system of equations for the gas component consists of the continuity equation, the momentum equation, and the energy balance equation. The gas dynamics is determined by the gravity of the central star and the self-gravity of the disc, viscosity, and friction between gas and dust. The energy balance in the disc depends on viscous heating, radiative heating (including the radiation of the central star and background radiation), radiative cooling, and adiabatic work, which can either heat or cool the local medium. In all equations, the influence of the magnetized disc wind on the gas is also taken into account. The corresponding equations in the thin-disc limit have the following form:
\begin{equation}
\frac{\partial \Sigma_{\rm g}}{\partial t} + \boldsymbol{\nabla} \cdot \left(\Sigma_{\rm g} \boldsymbol{v} \right) = -\dot{\Sigma}_{\rm w},
\label{eq: continuity-gas}
\end{equation}
\begin{equation}
\begin{aligned}
\frac{\partial \left(\Sigma_{\rm g} \boldsymbol{v} \right)}{\partial t} + \boldsymbol{\nabla} \cdot \left(\Sigma_{\rm g} \boldsymbol{v} \otimes \boldsymbol{v} \right) 
&= -\boldsymbol{\nabla} \mathcal{P} + \Sigma_{\rm g} \boldsymbol{g} + \boldsymbol{\nabla} \cdot \boldsymbol{\Pi}  \\
&\quad - \Sigma_{\rm d,gr} \boldsymbol{f} - \dot{\Sigma}_{\rm w} \boldsymbol{v} - \hat{\boldsymbol{e}}_\phi {T}_{z\phi}^{\rm w},
\end{aligned}
\label{eq: momentum-gas}
\end{equation}
\begin{equation}
\frac{\partial e}{\partial t} + \boldsymbol{\nabla} \cdot \left(e \boldsymbol{v} \right) = -\mathcal{P} \left(\boldsymbol{\nabla} \cdot  \boldsymbol{v} \right) - \Lambda + \Gamma + \boldsymbol{\nabla} \boldsymbol{v} : \Pi -\dot{\Sigma}_{\rm w} \frac{e}{\Sigma_{\rm g}}.
\label{eq: energy-gas}
\end{equation}
In the equations written above, $\Sigma_{\rm g}$ is the gas surface density, $e$ is the internal energy per unit surface area, $\boldsymbol{v} = v_{r} \hat{\boldsymbol{e}}_{r} + v_{\phi} \hat{\boldsymbol{e}}_\phi$ is 
the gas velocity in the disc plane, $\mathcal{P} = (\gamma_{\rm ad} - 1) e$ is the vertically integrated pressure obtained from the ideal gas equation of state (adiabatic index $\gamma_{\rm ad} = 7/5$), and $\boldsymbol{f}$ is the drag force (per unit mass) between gas and dust.
The loss of the total angular momentum per unit disc area due to the wind is characterized by the vertically integrated $\phi$-component of the divergence of the Maxwell stress tensor $\int (\boldsymbol{\nabla} \cdot \mathbf{\mathbf{T}}^{\rm w})_\phi \, dz$. Under our simplifying thin-disc assumptions, the integration yields the $z\phi$-component of the stress tensor at the disc surface $T_{z \phi}^{\rm w}$. The unit vector $\hat{\boldsymbol{e}}_\phi$ means that the magnetic disc wind acts only on the $\phi$-component of the momentum $\Sigma_{\rm g} \boldsymbol{v}$ in our model.

The gravitational acceleration in the disc plane: $\boldsymbol{g} = g_{r} \hat{\boldsymbol{e}}_r + g_{\phi} \hat{\boldsymbol{e}}_\phi$, accounts for the gravity of the central protostar after its formation, as well as the self-gravity of the gas and dust disc components. The total gravitational potential of gas and dust is determined by solving its integral form using the convolution method, as described in \citet{Binney1987}:
\begin{equation}
    \Phi(r, \phi) = -G \int^{r_{\rm out}}_{r_{\rm in}} r' dr' \int^{2 \pi}_{0} \frac{\left(\Sigma_{\rm g}(r', \phi') + \Sigma_{\rm d,tot}(r', \phi') \right) d\phi'}{\sqrt{(r')^2 + r^2 - 2 r r' \cos(\phi'-\phi)}},
\label{eq: grav_potential}
\end{equation}
where $r_{\rm in}$ and $r_{\rm out}$ are the radial distances to the inner and outer boundaries of the computational domain, $\Sigma_{\rm d,tot}$ is the total dust mass, and $G$ is the gravitational constant. Note that the convolution method does not necessarily require the introduction of a smoothing parameter to prevent the singularity when $r=r'$ and $\phi=\phi'$. Details of the potential calculation method without smoothing, test problems, and a comparison with the method using an explicit smoothing term are given in \citet{Vorobyov2024}. In Appendix~\ref{app: grav_pot_calc} of the present work, we also demonstrate that the disc evolution simulated without using a smoothing parameter does not differ significantly from the disc evolution in a model where a smoothing parameter is introduced.

The action of turbulent viscosity is represented by the viscous stress tensor, the expression for which has the form:
\begin{equation}
    {\boldsymbol{\Pi}} = 2 \Sigma \nu \left(\boldsymbol{\nabla v} - \frac{1}{3}(\boldsymbol{v} \cdot \boldsymbol{\nabla}) \mathbb{I} \right),
\label{eq: visc_stress_tensor}
\end{equation}
where $\mathbb{I}$ is the unit tensor, and $\nu$ is the kinematic viscosity.
The $\alpha$-parameterization \citep{ShakuraSunyaev1973} of the kinematic viscosity is used:
\begin{equation}
    \nu = \alpha_{\rm visc} c_{\rm s} H_{\rm g},
\label{eq: viscosity}
\end{equation}
where the parameter $\alpha_{\rm visc} = 10^{-3}$, $c_{\rm s}$ is the sound speed, and $H_{\rm g}$ is the gas vertical scale height, which is calculated taking into account the disc self-gravity.
Expressions for the cooling rates due to thermal dust emission $\Lambda$ and radiative heating (by the star and background interstellar radiation) $\Gamma$ can be found in \citet{Vorobyov2018}.

The model of the magnetic disc wind considers the case when the disc is threaded by a large-scale magnetic field. In the region between the disc surface and the Alfvén surface, where the magnetic pressure exceeds the ram pressure of the gas, the magnetic field lines are straight and rotate rigidly; therefore, the angular velocity of the matter in the disc $\Omega$ and the angular velocity at the Alfvén surface $\Omega_{\rm A}$ are equal. If the magnetic field lines are sufficiently inclined to the disc normal, matter can start moving along the line when the projection of the centrifugal force onto the field line exceeds the projection of the gravitational force. Moving along the line, the matter accelerates outward and increases its angular momentum. The rate of angular momentum loss from the Alfvén surface:
\begin{equation}
\dot{L}_{\rm w} = -r_{\rm A}^2 \Omega \dot{\Sigma}_{\rm w},
\label{eq: ang_mom_loss_wind-alfven}
\end{equation}
where $\dot{\Sigma}_{\rm w}$ is the rate of surface density loss due to the magnetic disc wind, $r_{\rm A}$ is the Alfvén radius, which can be expressed through the magnetic lever arm parameter $\lambda$, which defines the ratio of the specific angular momentum at the Alfvén surface to the specific angular momentum in the disc on the same magnetic field line:
\begin{equation}
\lambda = \frac{r_{\rm A}^2 \Omega_{\rm A}}{r^2 \Omega} = \left( \frac{r_{\rm A}}{r} \right)^2,
\label{eq: magn_lever_arm}
\end{equation}
where it is accounted for $\Omega_{\rm A} = \Omega$. The value $\lambda = 2$ is used in this work. Expressing $r_{\rm A}$ from the last equality and substituting it into equation (\ref{eq: ang_mom_loss_wind-alfven}), we obtain:
\begin{equation}
\dot{L}_{\rm w} = -\lambda r^2 \Omega \dot{\Sigma}_{\rm w}.
\label{eq: ang_mom_loss_wind-with-lambda}
\end{equation}
From this relation, it follows that the rate of total angular momentum loss per unit disc surface area can be represented in the following form:
\begin{equation}
\frac{d L_z}{dt} = -\lambda r v_{\phi} \dot{\Sigma}_{\rm w},
\label{eq: ang_mom_loss_wind_disk}
\end{equation}
On the other hand, the influence of the magnetic disc wind on the evolution of the disc angular momentum can be represented as the combined action of two processes: the removal of the disc's angular momentum together with the mass, and the braking of the disc due to the action of the magnetic torque:
\begin{equation}
\frac{d L_z}{dt} \equiv \frac{d}{dt} \left( \Sigma_{\rm g} r v_{\phi} \right) = -\dot{\Sigma}_{\rm w} r v_{\phi} - r T_{z \phi}^{\rm w}.
\label{eq: ang_mom_loss_wind}
\end{equation}
Following \citet{Tabone2022}, by analogy with kinematic viscosity, we define the $z\phi$-component of the stress tensor at the disc surface present in equation (\ref{eq: momentum-gas}) by introducing the parameter
\begin{equation}
\alpha_{\rm wind} \equiv \frac{4}{3} \frac{r T_{z \phi}^{\rm w}}{\Sigma_{\rm g} c_{\rm s}^2}.
\label{eq: alpha_wind}
\end{equation}
By equating the right-hand sides of equations (\ref{eq: ang_mom_loss_wind_disk}) and (\ref{eq: ang_mom_loss_wind}), and taking into account the definition (\ref{eq: alpha_wind}), the following expression for the mass loss rate due to the magnetic disc wind can be obtained:
\begin{equation}
   \dot{\Sigma}_{\rm w} = \frac{3 \alpha_{\rm wind} c_{\rm s}^2}{4 (\lambda - 1) \Omega r^2} \Sigma_{\rm g}.
\label{eq: sigma-dot-wind}
\end{equation}

We also note that a comparison of the accretion rates onto the star and the wind mass loss rates (see Appendix~\ref{app: accretion_rates}) confirms the fulfillment of basic energy principles (accretion-powered winds) and indicates the internal consistency of the model used.

\subsection{Dust component}
\label{Subsection: dust_component}

The dust component is divided into two populations:
\begin{enumerate}
    \item small dust, representing dust particles with sizes ranging from $a_{\rm min} = 5 \times 10^{-3}~{\rm \mu m}$ to $a_{*} = 1~{\rm \mu m}$;
    \item grown dust, the size of which varies from $a_{*}$ to the maximum value $a_{\rm max}$, the magnitude of which changes in space and time.
\end{enumerate}
Initially, all dust in the collapsing prestellar cloud belongs to the small dust population. During the disc formation and evolution, the small dust can grow and transform into grown dust. It is assumed that the dust in both populations is distributed in size according to a simple power law:
\begin{equation}
   N(a) = C \cdot a^{\rm -p},
\label{eq: dust-distribution}
\end{equation}
where $C$ is a normalization constant, and the power-law index ${\rm p} = 3.5$ does not change over time.

The continuity equations are solved separately for the grown and small dust ensembles, but the momentum equation is solved only for the grown dust, since small dust is assumed to be dynamically coupled to the gas. The system of hydrodynamic equations for dust in the zero-pressure limit is written as:
\begin{equation}
\frac{\partial \Sigma_{\rm d,sm}}{\partial t} + \boldsymbol{\nabla} \cdot \left(\Sigma_{\rm d,sm} \boldsymbol{v} \right) = -S(a_{\rm max}) -\dot{\Sigma}_{\rm w} \frac{\Sigma_{\rm d,sm}}{\Sigma_{\rm g}},
\label{eq: continuity-small-dust}
\end{equation}
\begin{equation}
\frac{\partial \Sigma_{\rm d,gr}}{\partial t} + \boldsymbol{\nabla} \cdot \left(\Sigma_{\rm d,gr} \boldsymbol{u} \right) = S(a_{\rm max}),
\label{eq: continuity-grown-dust}
\end{equation}
\begin{equation}
\begin{aligned}
\frac{\partial \left(\Sigma_{\rm d,gr} \boldsymbol{u} \right)}{\partial t} + \boldsymbol{\nabla} \cdot \left(\Sigma_{\rm d,gr} \boldsymbol{u} \otimes \boldsymbol{u} \right) 
&= \Sigma_{\rm d,gr} \boldsymbol{g} \\
&\quad + \Sigma_{\rm d,gr} \boldsymbol{f} + S(a_{\rm max})\boldsymbol{v},
\end{aligned}
\label{eq: momentum-dust}
\end{equation}
where $\Sigma_{\rm d,sm}$ is the surface density of small dust, $\Sigma_{\rm d,gr}$ is the surface density of grown dust, $\boldsymbol{u}$ is the velocity of grown dust in the disc plane, and $S$ is the conversion rate of small dust into grown dust (per unit disc surface area).
The mutual influence of gas and grown dust on each other is taken into account in equations (\ref{eq: momentum-gas}) and (\ref{eq: momentum-dust}) by including the term $\Sigma_{\rm d,gr} \boldsymbol{f}$, where $\boldsymbol{f}$ is the drag force (per unit mass) between gas and grown dust, which, depending on local conditions and dust properties, can be considered in two regimes: Epstein and Stokes (linear and nonlinear, respectively). We note, however, that in this work, in all the models used, the conditions did not go beyond the Epstein regime. The expression for $S(a_{\rm max})$ and a more detailed description of the approach to calculating the drag force can be found in \citet{Vorobyov2023} and references therein.

The evolution of the maximum dust size is described by the following equation:
\begin{equation}
\frac{\partial a_{\rm max}}{\partial t} + \left( \boldsymbol{u} \cdot \boldsymbol{\nabla} \right) a_{\rm max} = \mathcal{D},
\label{eq: a_max_evolution}
\end{equation}
where the dust growth rate due to collisions and coagulation is calculated in the monodisperse approximation \citep{Birnstiel2012}:
\begin{equation}
\mathcal{D} = \frac{\rho_{\rm d}}{\rho_{\rm s}}u_{\rm rel}.
\label{eq: dust_growth_rate}
\end{equation}
Here, $\rho_{\rm d}$ is the volume density of dust (both small and grown), $\rho_{\rm s} = 2.24~{\rm g \ cm^{-3}}$ is the material density of dust grains \citep{Weingartner2001}, and the dust-to-dust collision velocity is defined as $u_{\rm rel} = \sqrt{u_{\rm th}^{2} + u_{\rm turb}^{2}}$, where $u_{\rm th}$ and $u_{\rm turb}$ account for the Brownian and turbulence-induced local motion, respectively, and the latter is calculated following \citet{Ormel&Cuzzi2007} as
\begin{equation}
    u_{\rm turb} = \sqrt{\frac{3 \alpha_{\rm visc}}{{\rm St} + {\rm St}^{-1}} }  c_{\rm s},
\label{eq: u_turb}
\end{equation}
where ${\rm St}$ is the Stokes number corresponding to dust grains of maximum size $a_{\rm max}$.

Dust growth in our numerical model is limited by fragmentation and radial drift. The fragmentation barrier is defined, according to \citet{Birnstiel2016}, as
\begin{equation}
    a_{\rm frag} = \frac{2 \Sigma_{\rm g} v_{\rm frag}^2}{3 \pi \rho_{\rm s} \alpha_{\rm visc} c_{\rm s}^2},
\label{eq: fragmentation_barrier}
\end{equation}
and depends on the fragmentation velocity $v_{\rm frag}$ -- the threshold value of the relative velocity of dust particles at which collisions lead to fragmentation rather than coagulation. The drift barrier is taken into account self-consistently by calculating the dynamics of the grown dust. In Appendix~\ref{app: amax_evolution}, it is shown that the maximum dust size $a_{\rm max}$ in our model is in good agreement with analytical estimates of the dust growth and drift barriers $a_{\rm frag}$ and $a_{\rm drift}$, respectively.

The magnetized disc wind carries away only small dust in proportion to the local mass ratio of small dust to gas, therefore the continuity equation for small dust (\ref{eq: continuity-small-dust}) contains an additional term on the right-hand side. Grown dust in this model is not affected by the wind.

\subsection{Initial and boundary conditions}
\label{Subsection: init_bound_cond}

The numerical simulations start with the gravitational collapse of a prestellar core, which, in the adopted thin-disc approximation, has the shape of a flattened pseudo-disc. Such a spatial configuration can be expected in the presence of rotation and large-scale magnetic fields (e.g., \citet{Basu1997}). As the collapse proceeds, the inner regions of the core spin up, and when the inner infalling layers of the core encounter the centrifugal barrier near the inner boundary of the computational domain, a circumstellar disc is formed. The matter that has passed through the inner boundary of the computational domain before the formation of the circumstellar disc represents the embryo of the central star, which subsequently continues to grow due to accretion from the surrounding disc. Material infalling from the collapsing cloud core continues to arrive at the outer edge of the circumstellar disc until the core is exhausted. The rate of material infall onto the disc is consistent with what can be expected from analytical models of the collapse of spherical clouds \citep{Vorobyov2010}.

\begin{figure*}
    \centering
    \includegraphics[scale=0.45]{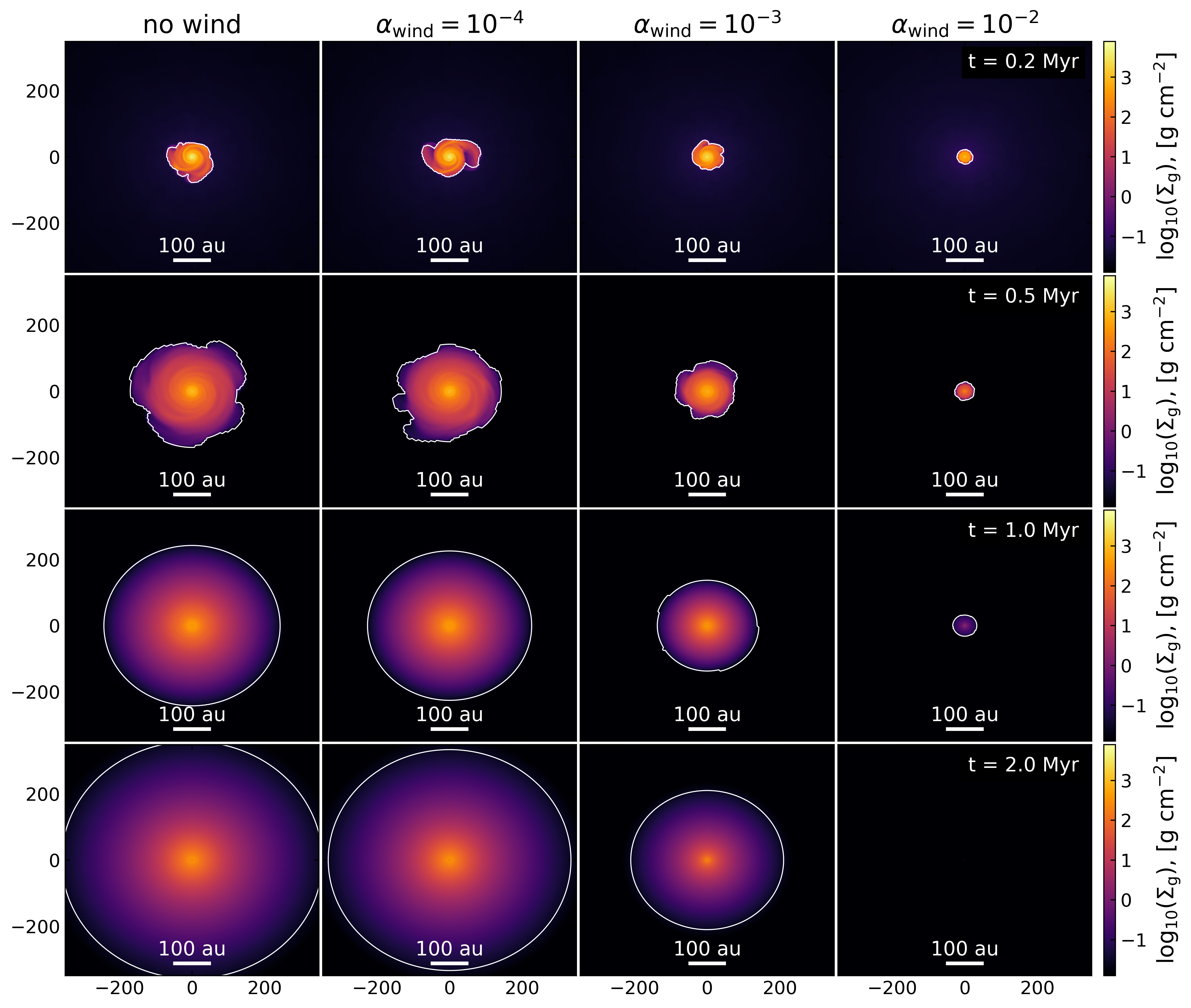}
    \caption{Two-dimensional distribution of the gas surface density in the disc at time instances (from top to bottom): 0.2 Myr, 0.5 Myr, 1 Myr, 2 Myr. The white curve indicates the disc gas radius.}
    \label{fig: 2D_gas_density}
\end{figure*}

The initial mass of the core in each model is $M_{\rm c} = 1~M_{\odot}$. 
The rotation rate of the core is set by specifying the ratio of rotational-to-gravitational energy $\beta \approx 10^{-3}$. This value lies within the limits inferred from observations of prestellar cloud cores, but is lower than the mean value of $\sim 2 \times 10^{-2}$ reported by \citet{Caselli2002}. 
A lower than average value is chosen to avoid strong gravitational instability and fragmentation in the disc. This branch of protoplanetary disc evolution deserves a separate  focused study, but requires a higher numerical resolution than adopted in this work.
Distributions for prestellar cores that are typical of vertically integrated Bonnor-Ebert spheres are used as initial surface density and angular velocity profiles \citep{Dapp2009}:
\begin{equation}
\Sigma_{\rm g}(r) = \frac{r_0 \Sigma_0}{\sqrt{r^2 + r_0^2}},
\label{eq: sigma_init}
\end{equation}
\begin{equation}
\Omega(r) = 2\Omega_0 \left( \frac{r_0}{r} \right)^2 \left[ \sqrt{1 + \left(\frac{r}{r_0} \right)^2} - 1 \right],
\label{eq: omega_init}
\end{equation}
where $\Sigma_0$ and $\Omega_0$ are the surface density and angular velocity at the core centre, respectively; and $r_0 = 1543~{\rm au}$ is the radius of the central plateau in the initial core.
The total dust-to-gas mass ratio is set to the interstellar medium value, $\xi_{\rm d2g} = (\Sigma_{\rm d,sm} + \Sigma_{\rm d,gr}) / \Sigma_{\rm g} = 10^{-2}$. The core and the subsequently formed disc are heated by background radiation with a temperature $T_{\rm bg} = 15~{\rm K}$, which is also taken as the initial temperature of the cloud.

The region between the inner computational boundary at $r_{\rm in} = 1~{\rm au}$ and the star is replaced with a sink cell, which allows free mass exchange (both inflow and outflow) across the sink-disc interface (see \citet{Vorobyov2018} for details). The outer boundary is open to mass outflow, but mass inflow from outside the computational domain is not permitted.
The computations continue up to 2 Myr, i.e., over timescales significantly longer than the dissipation time of the parent core.

\section{RESULTS}
\label{Section: results}

\begin{table}
    \caption{The parameters of the models used in simulations.}
    \label{tab: model_parameters}
    \begin{tabular}{lccccccc}
        \hline
        Section & $\alpha_{\rm wind}$ & $\alpha_{\rm visc}$ & $\lambda$ & $M_{\rm c}$ & $\xi_{\rm d2g}^{\rm init}$ & $v_{\rm frag}$\\
        & & & & $[M_{\odot}]$ & & ${[\rm m~s^{-1}]}$\\
        \hline
        \ref{Subsection: general_evolution}--\ref{Subsection: torques} & no wind & $10^{-3}$ & $2$ & $1.0$ & $10^{-2}$ & $5.0$\\
        \ref{Subsection: general_evolution}--\ref{Subsection: torques} & $10^{-4}$ & $10^{-3}$ & $2$ & $1.0$ & $10^{-2}$ & $5.0$\\
        \ref{Subsection: general_evolution}--\ref{Subsection: torques} & $10^{-3}$ & $10^{-3}$ & $2$ & $1.0$ & $10^{-2}$ & $5.0$\\
        \ref{Subsection: general_evolution}--\ref{Subsection: torques} & $10^{-2}$ & $10^{-3}$ & $2$ & $1.0$ & $10^{-2}$ & $5.0$\\
        \hline
        \ref{Subsection: dust_depletion} & $10^{-3}$ & $10^{-3}$ & $2$ & $1.0$ & $10^{-2}$ & $5.0$\\
        \ref{Subsection: dust_depletion} & $10^{-3}$ & $10^{-3}$ & $2$ & $1.0$ & $10^{-2}$ & $1.0$\\
        \ref{Subsection: dust_depletion} & $10^{-3}$ & $10^{-3}$ & $2$ & $1.0$ & $10^{-2}$ & $0.5$\\
        \hline
    \end{tabular}
\end{table}

This section presents the results of numerical simulations of protoplanetary disc evolution up to $2$ Myr for three models with winds of varying intensity, $\alpha_{\rm wind} = 10^{-4}$, $\alpha_{\rm wind} = 10^{-3}$, $\alpha_{\rm wind} = 10^{-2}$, as well as for a fiducial model (no wind), in which disc evolution occurs solely through self-gravity and viscosity. In all models, $\alpha_{\rm visc} = 10^{-3}$.
Section~\ref{Subsection: general_evolution} describes the global picture of protoplanetary disc evolution for the listed models. 
The analysis of the masses and sizes of protoplanetary discs is presented in Section~\ref{Subsection: masses_and_sizes}. 
To explain the differences in the evolution of discs with and without a magnetic disc wind, a study of magnetic, gravitational, and viscous torques was carried out, the results of which are presented in Section~\ref{Subsection: torques}. 
Section~\ref{Subsection: dust_depletion} separately considers the evolution of models with $\alpha_{\rm wind} = 10^{-3}$, but with different values of the fragmentation velocity (in addition to the reference value of $5~{\rm m~s^{-1}}$ used in the previous sections, values of $1~{\rm m~s^{-1}}$ and $0.5~{\rm m~s^{-1}}$ are also considered). The parameters of the models used in our simulations are summarised in Table~\ref{tab: model_parameters}.

\subsection{General evolution}
\label{Subsection: general_evolution}

The initial stage of evolution, from $t = 0$, which in this work corresponds to the onset of the gravitational collapse of the molecular cloud, to $t = 0.09$ Myr, corresponding to the formation of a circumstellar disc, is practically identical in all models, since the influence of the magnetic disc wind is limited to the extent of the disc and is negligible in the envelope. The subsequent evolution of the formed disc is of interest.

Fig.~\ref{fig: 2D_gas_density} shows the two-dimensional distribution of the gas surface density in the disc at different evolutionary stages, starting from $0.2$ Myr. The white curve denotes the outer boundary of the \textit{gas} disc.
The definition of the gas disc radius in our numerical model is based on two criteria \citep{Kadam2025}:
\begin{enumerate}
    \item Gas is considered to be part of the disc if it is located at such a radial distance $r$ that the magnitude of the gravitational acceleration of that gas element,
    \begin{equation}
        a_{\rm gr} = \frac{GM_{\rm enc}}{r^2},
    \label{eq: grav_acceleration}
    \end{equation}
   does not exceed the magnitude of the centrifugal acceleration,
    \begin{equation}
        a_{\rm cf} = \frac{v_{\phi}^2}{r},
    \label{eq: cf_acceleration}
    \end{equation}
     where $M_{\rm enc}$ is the mass enclosed within an orbit of radius $r$. This condition can be written as $a_{\rm gr} \le \zeta_{\rm crit} a_{\rm cf}$, where the parameter $\zeta_{\rm crit} = 1.5$, since matter in the disc moves at a velocity slightly different from Keplerian due to the contribution of the disc's self-gravity. Taking into account equations (\ref{eq: grav_acceleration}) and (\ref{eq: cf_acceleration}), the first criterion takes the following form:
    \begin{equation}
       \frac{GM_{\rm enc}}{v_{\phi}^2 r} \le \zeta_{\rm crit}.
    \label{eq: first_criterion_1}
    \end{equation}
    The distance in the disc at which infalling material with a specific angular momentum $j = v_{\phi} r$ crosses the disc plane is called the centrifugal radius:
    \begin{equation}
        r_{\rm cf} = \frac{j^2}{GM_{\rm enc}}.
    \label{eq: centrifugal_radius}
    \end{equation}
    The criterion (\ref{eq: first_criterion_1}) can be viewed as a constraint on the centrifugal radius, and it can be recast in more compact form:
    \begin{equation}
       \frac{r}{r_{\rm cf}} \le \zeta_{\rm crit}.
    \label{eq: first_criterion_2}
    \end{equation}
    
    \item The second criterion limits the disc size by a critical density value:
    \begin{equation}
        \Sigma_{\rm g} \ge \Sigma_{\rm crit},
    \label{eq: second_criterion_Rgas}
    \end{equation}
    where the value $\Sigma_{\rm crit} = 3 \times 10^{-2}\ {\rm g\ cm}^{-2}$ is chosen to avoid overestimating the gas disc radius by the first criterion at late times, when the infall from the cloud core ceases and viscous disc expansion becomes significant.
\end{enumerate}
The gas disc radius $R_{\rm gas}$ is defined as the boundary of the region inside which both criteria (\ref{eq: first_criterion_2}) and (\ref{eq: second_criterion_Rgas}) are satisfied in every cell.
Inside this region, the magnetic disc wind is active, and outside it, it is suppressed, according to our numerical model. It can be seen that the disc radius increases with time in all models, but the stronger the wind, the smaller the radius. In the model with the strongest wind, $\alpha_{\rm wind} = 10^{-2}$, the disc expansion is barely detectable. Moreover, in this model, by the time $t \approx 1.25$ Myr, the disc becomes completely depleted, since the gas mass in the disc $M_{\rm gas} < 1\ M_{\oplus}$. Therefore, in this and other figures, plots for later evolutionary stages in this model are not shown.

\begin{figure}
    \centering
    \includegraphics[width=\columnwidth]{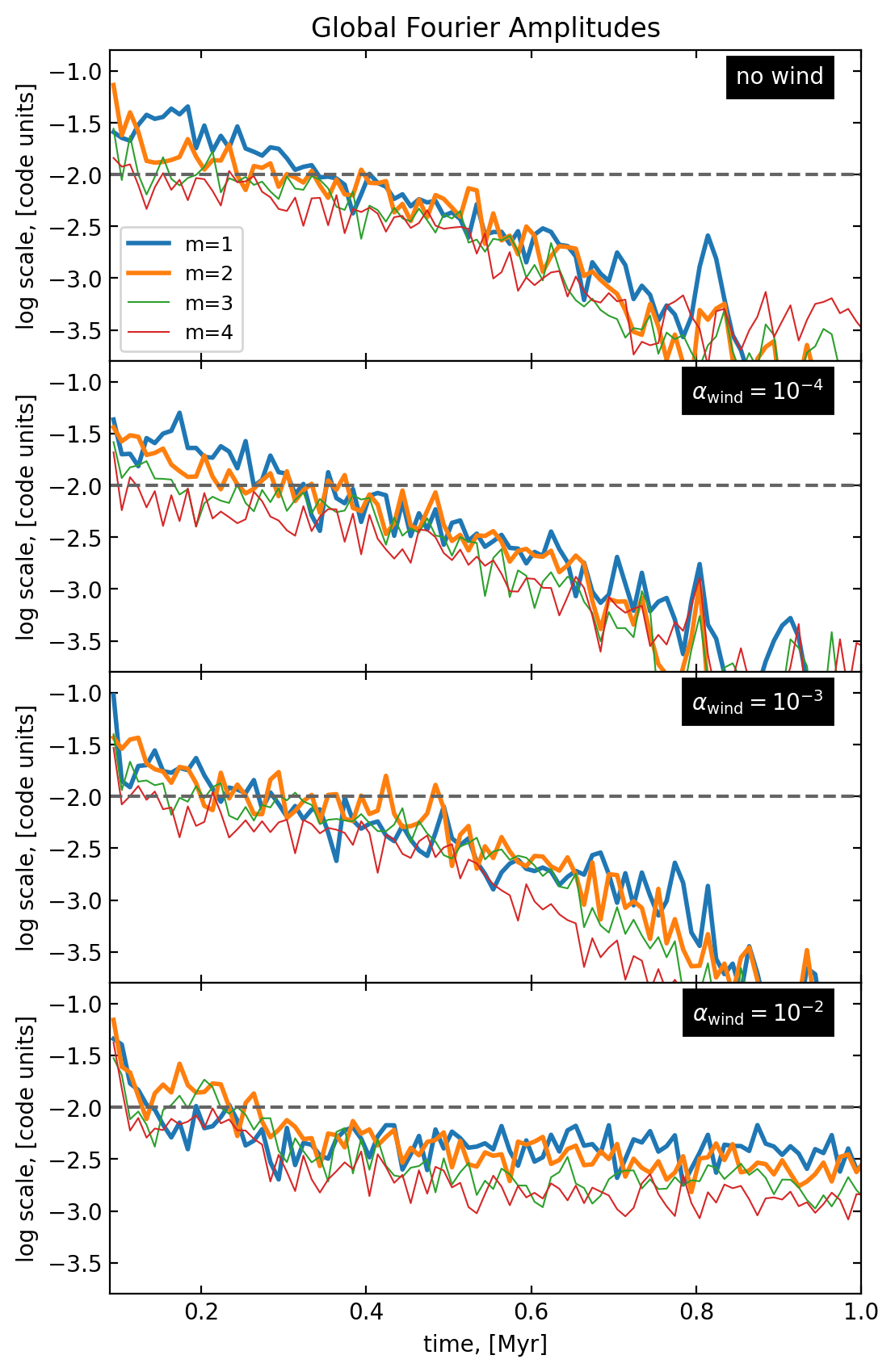}
    \caption{Evolution of the time-averaged (10 kyr) global Fourier amplitudes in the disc, representing spiral density wave perturbations in the disc relative to the axisymmetric density distribution. The dominant modes $m = 1$ and $m = 2$ are shown with a thicker lines.}
    \label{fig: global_fourier_amplitudes}
\end{figure}

\begin{figure*}
    \centering
    \includegraphics[scale=0.6]{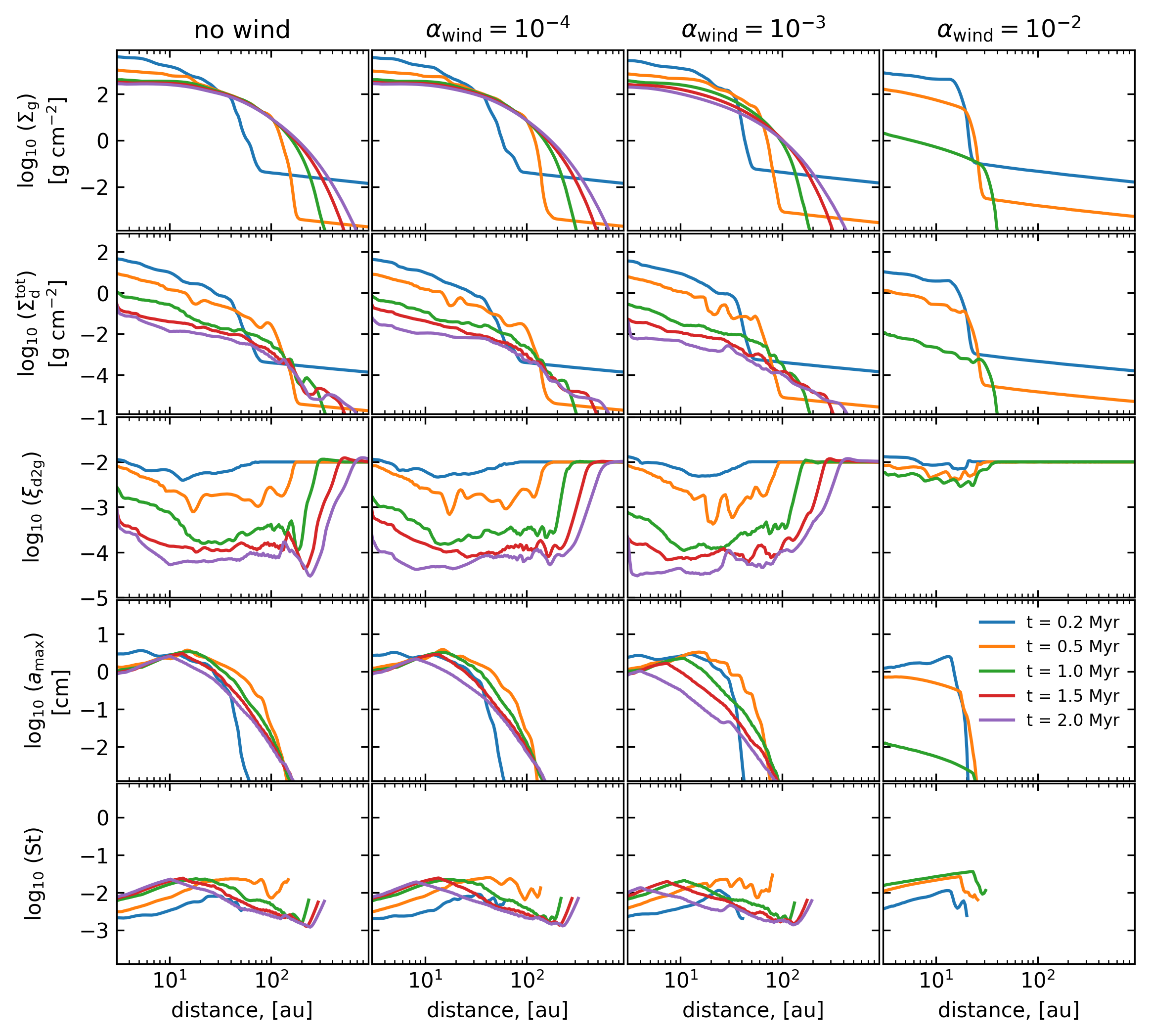}
    \caption{Evolution of radial profiles (from top to bottom): gas surface density, total dust surface density (sum of grown and small dust), total dust-to-gas mass ratio, maximum dust grain size, and Stokes number for dust. In the model with $\alpha_{\rm wind} = 10^{-2}$, the disc is completely depleted after 1.25 Myr and subsequent profiles are not shown. Radial profiles of the Stokes number are shown only within the disc radius.}
    \label{fig: density_profiles}
\end{figure*}

It can also be noted that for about half a million years, while material is infalling from the collapsing envelope, the disc exhibits a spiral structure, but subsequently becomes more axisymmetric. Interestingly, in the models with a wind, the spiral is less prominent. The analysis of the Fourier amplitudes, presented in Fig.~\ref{fig: global_fourier_amplitudes}, shows that the presence of a magnetic wind weakens gravitational instability at the initial stages of evolution. Fourier amplitudes can be viewed as a measure of the amplitude of spiral density wave perturbations in the disc compared to the axisymmetric density distribution. To calculate the global Fourier amplitudes in the disc as a function of time, the following formula was used:
\begin{equation}
    \left| A_{m}(t) \right| = \frac{1}{M_{\rm d} (t)}\left|\int\limits_{0}^{2\pi}{\int\limits_{3\ {\rm au}}^{R_{\rm gas}}{\Sigma_{\rm g}\left(r,\phi,t\right)}e^{im\phi}dr\ d\phi}\right|,
\label{eq: Fourier_amplitudes}
\end{equation}
where $M_{\rm d}(t)$ is the disc mass at a given time $t$, and $m$ is the spiral mode. The integral over distance is limited by the gas disc radius and starts from 3 au to avoid the influence of the inner boundary of the computational domain. From Fig.~\ref{fig: global_fourier_amplitudes}, it can be seen that after 0.4 Myr in all models, the amplitude values drop below $0.01$, but in the models with a more intense wind, starting from $\alpha_{\rm wind} = 10^{-3}$, a noticeable decrease in the magnitude of the dominant mode occurs earlier. It is worth noting that after 0.4 Myr in the model with $\alpha_{\rm wind} = 10^{-2}$, the Fourier amplitudes cease to change significantly, remaining within the range of $10^{-2}$ -- $10^{-3.5}$, while in the other models a continued decrease is observed. The explanation of this fact requires further study; however, such weak perturbations no longer have a significant impact on disc evolution via gravitational torques.

The top panel of Fig.~\ref{fig: density_profiles} shows the radial profiles of the azimuthally averaged gas surface density at different times. For the fiducial model (no wind), a decrease in the surface density in the inner disc regions and an increase in the disc size are observed during the first million years of evolution, while from 1 to 2 Myr, the change in the surface density in the inner regions is insignificant, whereas beyond 100 au the disc continues to expand gradually.
In the models with $\alpha_{\rm wind} = 10^{-4}$ and $\alpha_{\rm wind} = 10^{-3}$, at the corresponding times, one can notice slightly lower gas surface densities compared to the fiducial model and a less prominent expansion in the outer regions, while in the model with the intense wind, $\alpha_{\rm wind} = 10^{-2}$, a significant decrease in the gas surface density throughout the disc is already observed by 1 Myr of evolution, and disc expansion is almost completely suppressed.

The radial profiles of the azimuthally averaged total dust surface density $\Sigma_{\rm d, tot} = \Sigma_{\rm d,sm} + \Sigma_{\rm d,gr}$ and the dust-to-gas mass ratio $\xi_{\rm d2g}$ are shown in the second and third panels of Fig.~\ref{fig: density_profiles}, respectively. One can notice that even in the model without magnetic winds, by 2 Myr a significant decrease in the dust surface density occurs in the inner 100 au, which leads to a decrease in the $\xi_{\rm d2g}$ value by almost 2 orders of magnitude compared to the initial value of $0.01$. The models with winds $\alpha_{\rm wind} = 10^{-4}$ and $\alpha_{\rm wind} = 10^{-3}$ show a similar behavior of profile evolution, although the values of $\Sigma_{\rm d, tot}$ and $\xi_{\rm d2g}$ are slightly lower than in the fiducial model at the corresponding times. In the model with $\alpha_{\rm wind} = 10^{-2}$, a significant decrease in dust surface density is also observed, which, as in the case of gas, is more prominent than in the other models; however, the dust-to-gas ratio remains at the level of $10^{-2} - 10^{-2.5}$.
Since the decrease in the dust surface density is observed in all models, including the fiducial model (no wind), it can be concluded that the cause of this decrease is not the influence of the magnetic wind, as the wind entrails only small dust and does not carry away the mass-dominant grown dust. Therefore, dust depletion is caused by radial drift onto the star. In Appendix~\ref{app: char_drift_timescales}, it is shown that the spiral structure of the disc at the initial stages of evolution contributes to the temporary retention of dust, but nevertheless the characteristic drift timescales of grown dust are indeed comparable to the disc evolution time.
The less prominent decrease in the dust-to-gas ratio in the model with $\alpha_{\rm wind} = 10^{-2}$ is explained by the fact that the characteristic dust drift time in this model does not differ significantly from that in the other models, but the gas surface density decreases more intensively, which compensates for the drop in $\xi_{\rm d2g}$ due to dust drift.
We also note that lower values of the fragmentation velocity $v_{\rm frag}$ could prevent dust disc depletion. This case is considered separately in Section~\ref{Subsection: dust_depletion}.

The fourth and fifth panels of Fig.~\ref{fig: density_profiles} show the radial profiles of the azimuthally averaged maximum dust sizes and their Stokes numbers, respectively. The maximum dust size is limited by the fragmentation barrier $a_{\rm frag}$ and the radial drift barrier $a_{\rm drift}$. As shown in Appendix~\ref{app: amax_evolution}, in the inner disc regions the dust grain size is determined by the fragmentation barrier, while in the outer regions, especially at later times $t > 0.5$ Myr, the radial drift barrier becomes increasingly important. In this case, in the models with a more intense wind, the decrease in $a_{\rm max}$ is more prominent.

The dimensionless parameter, the Stokes number, is defined as the product of the stopping time and the Keplerian angular velocity: ${\rm St} = t_{\rm stop} \Omega_{\rm K}$. 
This definition is applicable within the disc, but not in the envelope, since the angular velocity there differs from Keplerian. Therefore, in Fig.~\ref{fig: density_profiles}, the radial profiles of this quantity beyond the outer radius of the disc are not shown.
The larger the Stokes number, the more time is required for the dust grain velocity to adapt to the gas velocity. Since $t_{\rm stop}$ depends linearly on the dust grain size (in the Epstein regime) and is inversely proportional to the gas density, the Stokes number is larger in those disc regions where larger dust grains reside and where the gas density is lower. Over time, the gas density in the inner region decreases, so an increase in the Stokes number is observed, while in the outer region a decrease occurs, since the disc expands and the gas density in this region becomes higher. As can be seen from Fig.~\ref{fig: density_profiles}, the Stokes number in the disc remains very low and does not even reach $10^{-1}$ throughout the entire considered time interval.

\subsection{Analysis of protoplanetary discs masses and sizes}
\label{Subsection: masses_and_sizes}

The gas disc mass $M_{\rm gas}$ is equal to the total gas mass from the inner disc boundary ($1~{\rm au}$) to the outer boundary $R_{\rm gas}$. The dust disc is considered to be the region within which 90\% of the total dust mass in the \textit{gas} disc is concentrated. This region is bounded by the radius $R_{\rm dust}$. This method of radius estimation is consistent with the approach to determining the masses and sizes of dust discs from observations, in which the radial distance bounding the region containing about 90\% of the flux is taken as the outer boundary \citep{Tobin2020}. Note, however, that mass is directly proportional to flux only in the optically thin case. Other factors like the instrument's sensitivity and resolution, together with the wavelength of observations and  uncertainties in opacities, can affect the one-to-one comparison as well \citep[e.g.,][]{Vorobyov2026}.

\begin{figure}
    \centering
    \includegraphics[width=\columnwidth]{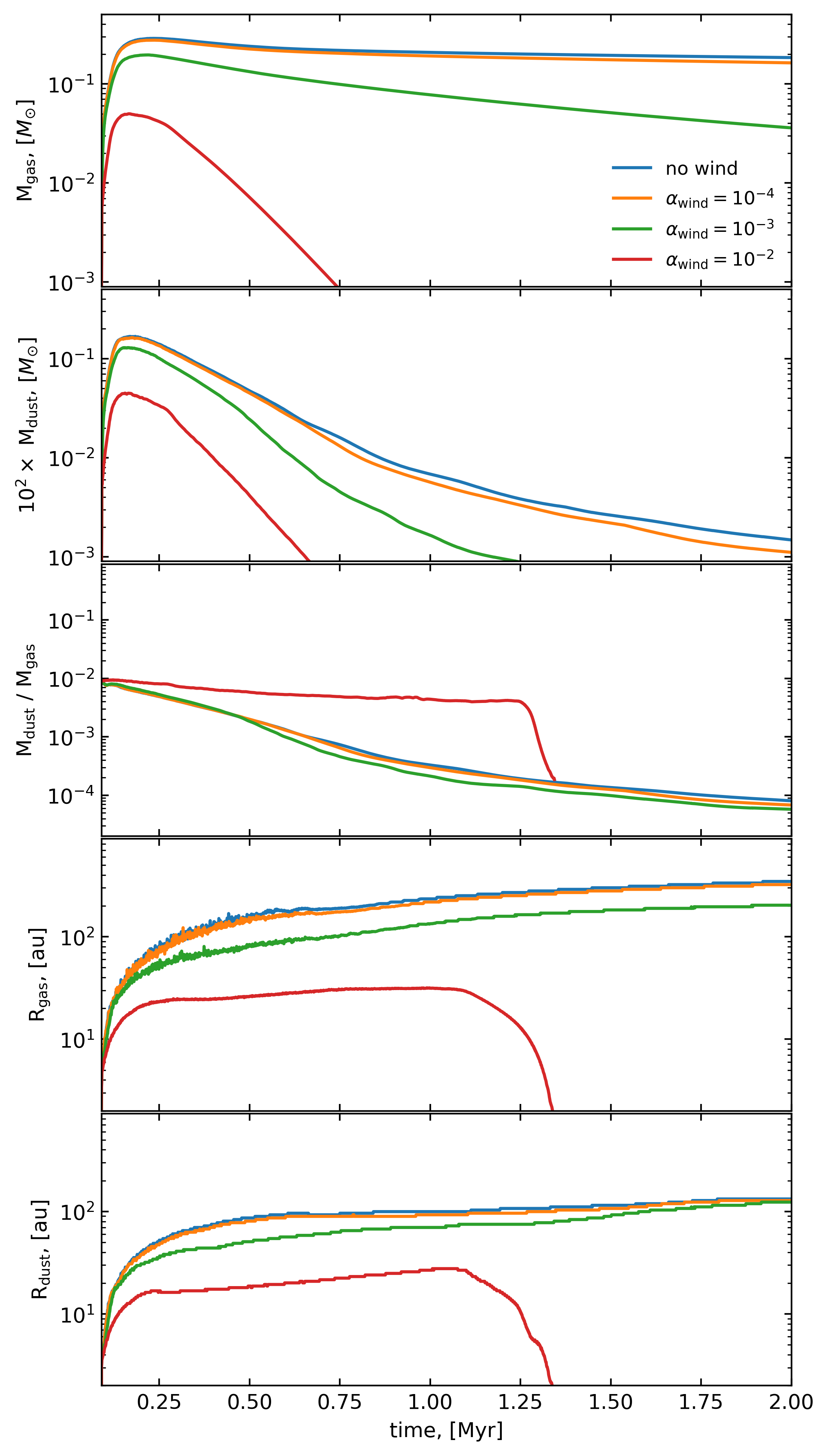}
    \caption{Evolution of the gas disc mass, dust disc mass, dust-to-gas mass ratio, gas disc radius, and dust disc radius (from top to bottom).} 
    \label{fig: gas_and_dust_radii_masses}
\end{figure}

The first and second panels of Fig.~\ref{fig: gas_and_dust_radii_masses} show the change over time of the gas and dust disc mass. Note that the second panel presents the dust disc mass in solar masses increased by a factor of 100 (for convenience of comparison with the gas disc mass). In all models, a brief period of mass growth at the initial stage of disc formation is quickly replaced by its decline. Moreover, the more intense the wind, the faster the decrease in mass occurs. The dust mass decreases significantly faster than the gas mass. This is confirmed by the third panel of Fig.~\ref{fig: gas_and_dust_radii_masses}, which shows the ratio $M_{\rm dust} / M_{\rm gas}$ of the dust disc mass to the gas disc mass. By the time $t = 2$ Myr, this ratio drops to $10^{-4}$ in all models, except for $\alpha_{\rm wind} = 10^{-2}$. In this model, the gas disc is depleted significantly faster compared to the other models. The ratio $M_{\rm dust} / M_{\rm gas}$ decreases slowly and remains at a level of $10^{-2} - 5 \times 10^{-3}$ because 
the rapid decline in $M_{\rm gas}$ (denominator) due to strong wind somewhat compensates for the rapid decline in  $M_{\rm dust}$ (numerator) due to fast dust drift. This trend persists until the final dissipation of the disc ($M_{\rm gas} < 1\ M_{\oplus}$), which occurs fairly rapidly around $t = 1.25$ Myr.

The fourth and fifth panels of Fig.~\ref{fig: gas_and_dust_radii_masses} demonstrate the evolution of the gas and dust disc sizes, respectively. Over time, the radii increase in all models, even in the model with $\alpha_{\rm wind} = 10^{-2}$, a slight growth is observed during the first million years. At late times, the gas disc size continues to increase due to viscous expansion, and a decrease in the radius is observed only in the model with the most intense wind. In the models with lower values of $\alpha_{\rm wind}$, external photoevaporation is likely required to limit the disc sizes.
Differences in the behavior of $R_{\rm gas}$ and $R_{\rm dust}$ are noticeable for the model with $\alpha_{\rm wind} = 10^{-3}$: while the difference between the final gas disc radius in this model (green curve) and in the fiducial (blue curve) or $\alpha_{\rm wind} = 10^{-4}$ (orange curve) model is noticeable, the final dust disc radius is almost the same in all three models. This can be explained by a significant decrease in the dust surface density in the inner disc by 2 Myr. Consequently, 90\% of the dust mass resides in a wide outer region (the more prominent the depletion, the closer to the gas disc boundary), where its large area compensates for the lower dust surface density. We note that other approaches to determining the dust disc radii may result in a different behavior of $R_{\rm dust}$ over time. As shown in Appendix~\ref{app: dust_disc_radii}, $R_{\rm dust}$ may decrease even in the model with no wind, if the dust disc radius is limited by a threshold value of $\Sigma_{\rm d, tot}^{\rm crit}$.

\begin{figure*}
    \centering
    \includegraphics[scale=0.66]{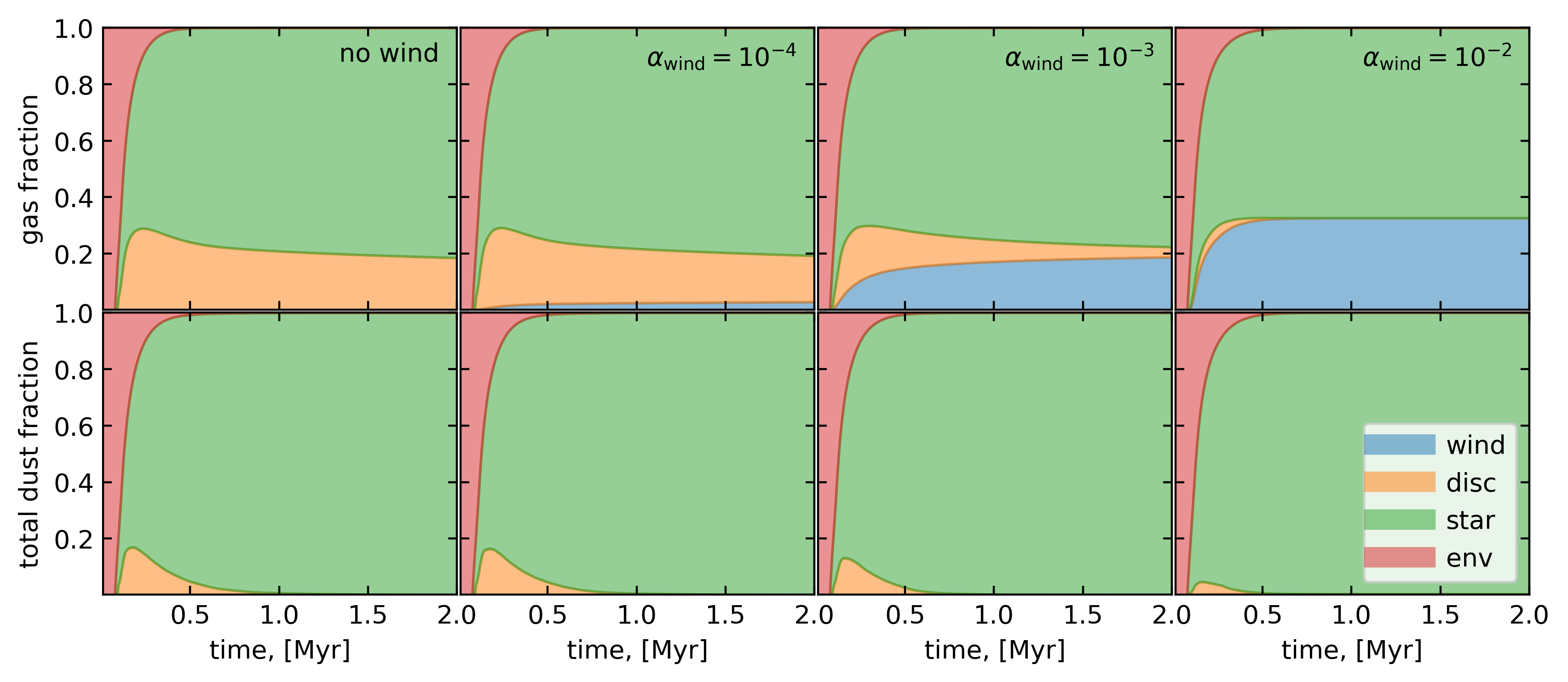}
    \caption{Evolution of the mass fraction of gas (top panel) and dust (sum of grown and small dust, bottom panel) contained in the envelope (red), on the star (green), in the disc (orange), and carried away by the wind (blue).}
    \label{fig: mass-gas-dust-fraction}
\end{figure*}

In order to better visualize the distribution of gas and dust mass, we refer to Fig.~\ref{fig: mass-gas-dust-fraction}, which shows in color the evolution of the fraction of gas mass (top panel) and dust mass (bottom panel). In all models, the mass initially contained in the envelope is transferred over time first to the disc and the star. Subsequently, in the case of gas mass, it can be seen that in models with a sufficiently strong wind, $\alpha_{\rm wind} = 10^{-3}$ and $\alpha_{\rm wind} = 10^{-2}$, a significant fraction (up to 30\% in the model with $\alpha_{\rm wind} = 10^{-2}$) of the mass leaves the system along with the magnetic disc wind. The dust mass, regardless of the model, eventually ends up on the star. The wind affects only small dust and carries away only an insignificant fraction of the total dust mass.   
Even in the model with $\alpha_{\rm wind} = 10^{-2}$ less than 1\% is lost by $t$ = 2~Myr. A sufficiently high value of the fragmentation velocity ($v_{\rm frag} = 5~{\rm m~s^{-1}}$), despite the temporary retention of dust by the spiral structure at the early stages, leads to the bulk of solid matter being driven to the star after one million years of evolution. However, note that the dynamics of dust in the inner region ($1~{\rm au}$) is not tracked. A fraction of the dust may be carried away by protostellar jets or locked in the inner dead zone with the possibility of turning into planetesimals due to the streaming instability \citep{Vorobyov2024,Cecil2024}.  Therefore, without an analysis of the processes inside $1~{\rm au}$, it is impossible to reliably state which fraction of heavy elements will end up on the star.

\subsection{Torques evolution}
\label{Subsection: torques}

The simulation results presented in Sections \ref{Subsection: general_evolution} and \ref{Subsection: masses_and_sizes} indicate that, while the evolutionary picture is similar in the fiducial model (no wind) and the model with a weak magnetic disc wind ($\alpha_{\rm wind} = 10^{-4}$), in the models with a sufficiently intense wind, starting from the model with $\alpha_{\rm wind} = 10^{-3}$, the differences in disc evolution become significant.
To understand the cause of these differences and to determine the dominant angular momentum transport mechanism among gravitational instability, viscosity, and the magnetic wind, we analyse the corresponding torques.
To do this, the local torques are calculated in each cell of the computational domain: the magnetic torque $\Gamma_{\rm wind}(r, \phi)$, the gravitational torque $\Gamma_{\rm grav}(r, \phi)$, and the viscous torque $\Gamma_{\rm visc}(r, \phi)$.

The formula for calculating the local magnetic torque is:
\begin{equation}
    \Gamma_{\rm wind}(r, \phi) = -r T_{z \phi}^{\rm w} S_{\rm cell}(r,\phi),
\label{eq: magn_torque_loc}
\end{equation}
where $S_{\rm cell}(r,\phi)$ is the cell area. The torque exerted by the disc wind per unit area, $rT_{z \phi}^{\rm w}$, can be expressed from the definition of the parameter $\alpha_{\rm wind}$ (\ref{eq: alpha_wind}), and, taking into account equations (\ref{eq: ang_mom_loss_wind-with-lambda}) and (\ref{eq: sigma-dot-wind}), can be explicitly related to the total angular momentum loss rate $\dot{L}_{\rm w}$ per unit disc area:
\begin{equation}
    -r T_{z \phi}^{\rm w} = \dot{\Sigma}_{\rm w} \Omega r^2 (\lambda - 1) = \frac{\lambda - 1}{\lambda} \dot{L}_{\rm w}.
\label{eq: T_z_phi}
\end{equation}
The local gravitational and viscous torques are calculated using the following classical formulas:
\begin{equation}
    \Gamma_{\rm grav}(r, \phi) = -m(r, \phi) \frac{\partial \Phi}{\partial \phi},
\label{eq: grav_torque_loc}
\end{equation}
\begin{equation}
    \Gamma_{\rm visc}(r, \phi) = r (\nabla \cdot {\bf{\Pi}})_{\phi} S_{\rm cell}(r,\phi),
\label{eq: visc_torque_loc}
\end{equation}
where $m(r, \phi)$ is the gas mass in the cell, $\Phi$ is the gravitational potential, and $\nabla \cdot {\bf{\Pi}}$ is the divergence of the viscous stress tensor, the $\phi$-component of which can be written as follows:
\begin{equation}
    (\nabla \cdot {\bf{\Pi}})_{\phi} = \frac{\partial}{\partial r} \Pi_{\phi r} + \frac{1}{r} \frac{\partial}{\partial \phi} \Pi_{\phi \phi} + \frac{2}{r} \Pi_{r \phi}.
\label{eq: nabla_Pi_phi}
\end{equation}

\begin{figure*}
    \centering
    \includegraphics[scale=0.28]{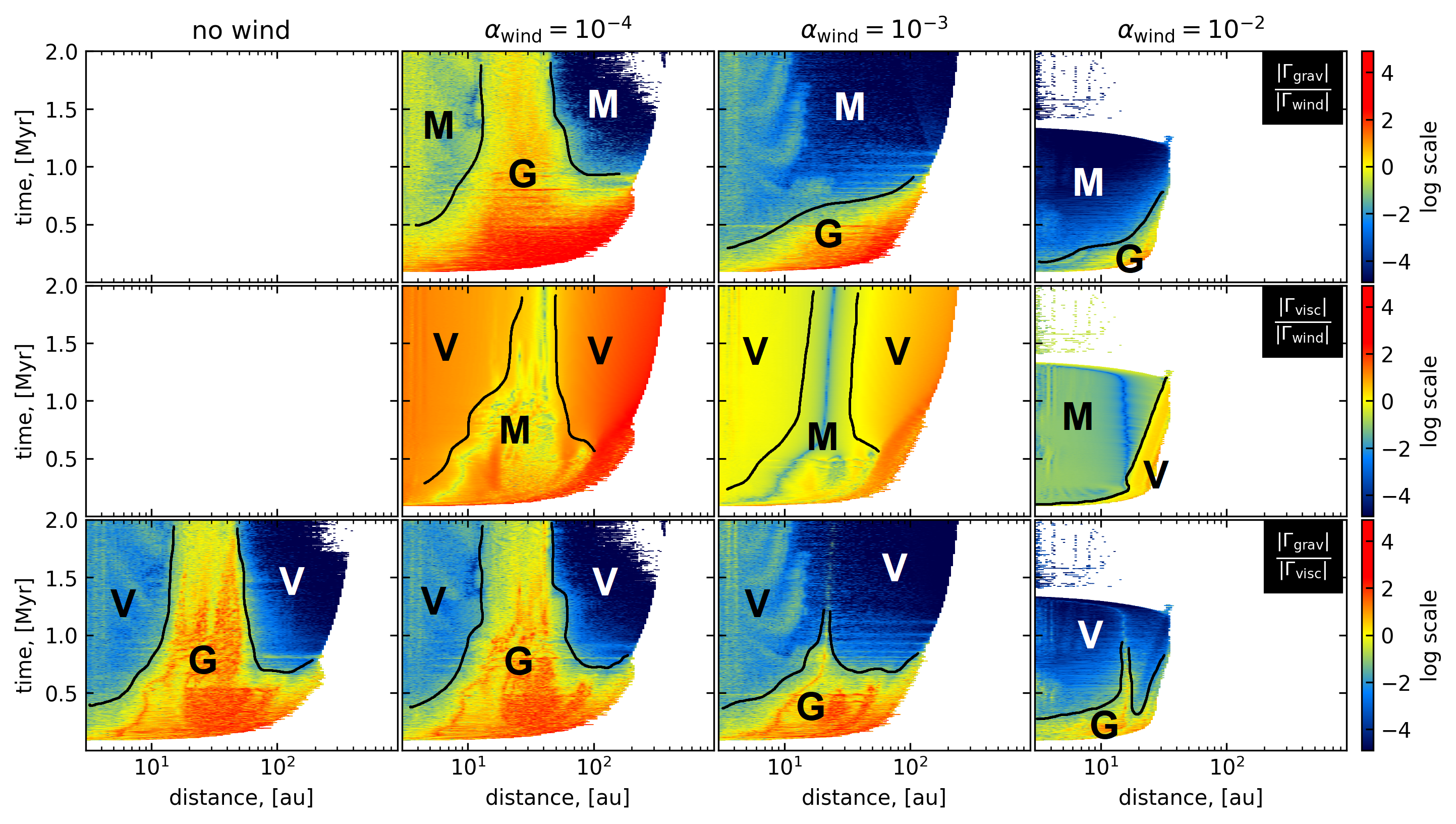}
    \caption{Temporal evolution of the ratio of the azimuthally averaged torques by absolute value (from top to bottom): gravitational to magnetic, viscous to magnetic, gravitational to viscous. The sign is taken into account when averaging local torques azimuthally. The letters mark the regions of prevalence of angular momentum transfer mechanisms: M -- magnetic torque, G -- gravitational torque, V -- viscous torque.}
    \label{fig: disk_torques_colormaps}
\end{figure*}

Fig.~\ref{fig: disk_torques_colormaps} shows the temporal evolution of the ratio of absolute values of azimuthally averaged torques (with the sign of the torque taken into account when averaging over azimuth): gravitational to magnetic -- top panel, viscous to magnetic -- middle panel, gravitational to viscous -- bottom panel. Yellow color in the figure means that the corresponding torques in a given region are equal and their ratio is 1, red color shows regions where the ratio is greater than 1, and blue -- less than 1. The approximate boundaries of the regions where one or another mechanism of angular momentum transfer prevails are also indicated. As can be seen, in almost all cases at the initial stages of evolution (until approximately 0.5 Myr), the gravitational torque dominates in most of the disc, then the viscous torque begins to prevail, especially in the outer regions. The only exception is the model with $\alpha_{\rm wind} = 10^{-2}$, where the gravitational and viscous torques generally do not exceed the magnetic torque in magnitude. In the middle and bottom panels of Fig.~\ref{fig: disk_torques_colormaps}, a region in the disc (located in most cases between 10 and 50 au) is clearly visible where a significant decrease in the magnitude of the viscous torque occurs. In Appendix~\ref{app: viscous_transport}, it is shown that in such regions the dynamic viscosity $\mu$ decreases with distance as $r^{-1/2}$, and the sign of the viscous torque changes from negative to positive, which explains the decrease in its absolute value. We also note that artificial viscosity has no effect on the disc  evolution in any of the models (see Appendix~\ref{app: artificial_viscosity}).

\begin{figure}
    \centering
    \includegraphics[width=\columnwidth]{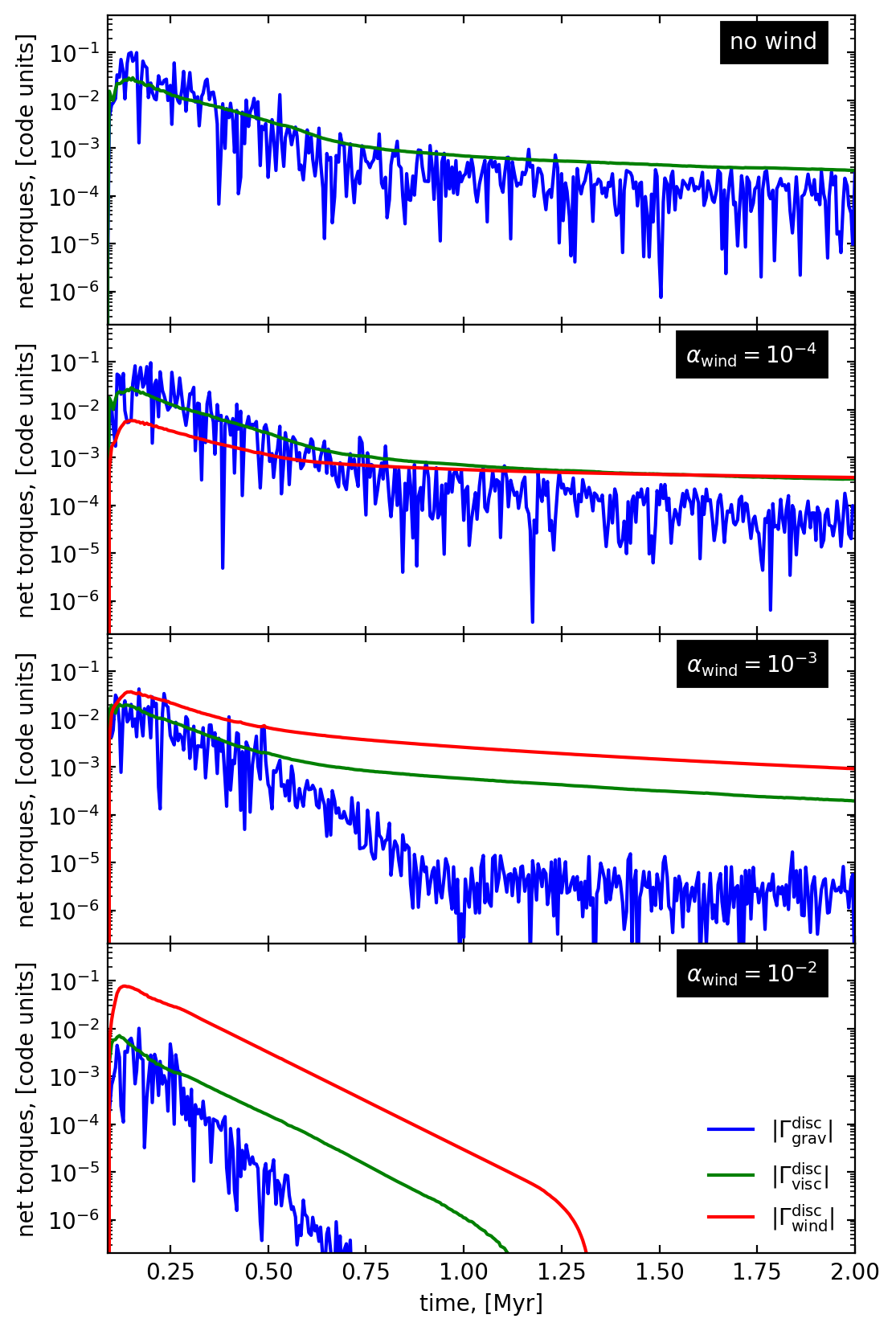}
    \caption{Evolution of the absolute magnitude of the net torques, averaged over a time interval of 10 kyr, in the models used. The curves show: gravitational torque (blue), viscous torque (green), and magnetic torque (red).}
    \label{fig: disk_torques}
\end{figure}

To analyse the influence of one or another mechanism on the disc evolution, we considered the change over time of net torques for the entire gas disc, presented in Fig.~\ref{fig: disk_torques}, which are calculated using the following formulas:
\begin{equation}
    \Gamma_{\rm wind}^{\rm disc} = \sum_{\phi} \sum_{r=3~{\rm au}}^{R_{\rm gas}} \Gamma_{\rm wind}(r,\phi),
\label{eq: net_magn_torq_disk}
\end{equation}
\begin{equation}
    \Gamma_{\rm grav}^{\rm disc} = \sum_{\phi} \sum_{r=3~{\rm au}}^{R_{\rm gas}} \Gamma_{\rm grav}(r,\phi),
\label{eq: net_grav_torq_disk}
\end{equation}
\begin{equation}
    \Gamma_{\rm visc}^{\rm disc} = \sum_{\phi} \sum_{r=3~{\rm au}}^{R_{\rm gas}} \Gamma_{\rm visc}(r,\phi).
\label{eq: net_visc_torq_disk}
\end{equation}
The dominance of gravity at the initial stage of evolution is replaced by the dominance of viscosity; however, in all models with the wind, the prevalence of the total magnetic torque is prominent, and it eventually becomes dominant even in the model with $\alpha_{\rm wind} = 10^{-4}$.
This can be explained by the fact that the local torques are summed with their signs taken into account. The magnetic torque is always negative, since it leads to a decrease in the total angular momentum, while the gravitational and viscous torques can be both positive and negative. As shown in Appendix~\ref{app: torques_colormaps}, when summing the local gravitational and viscous torques from the outer region to the inner one, their sum at some point starts to decrease, which indicates that in the inner part of the disc these torques are predominantly negative, while in the outer part they are positive (see also \citet{Vorobyov2007, Das2026}).

\begin{figure}
    \centering
    \includegraphics[width=\columnwidth]{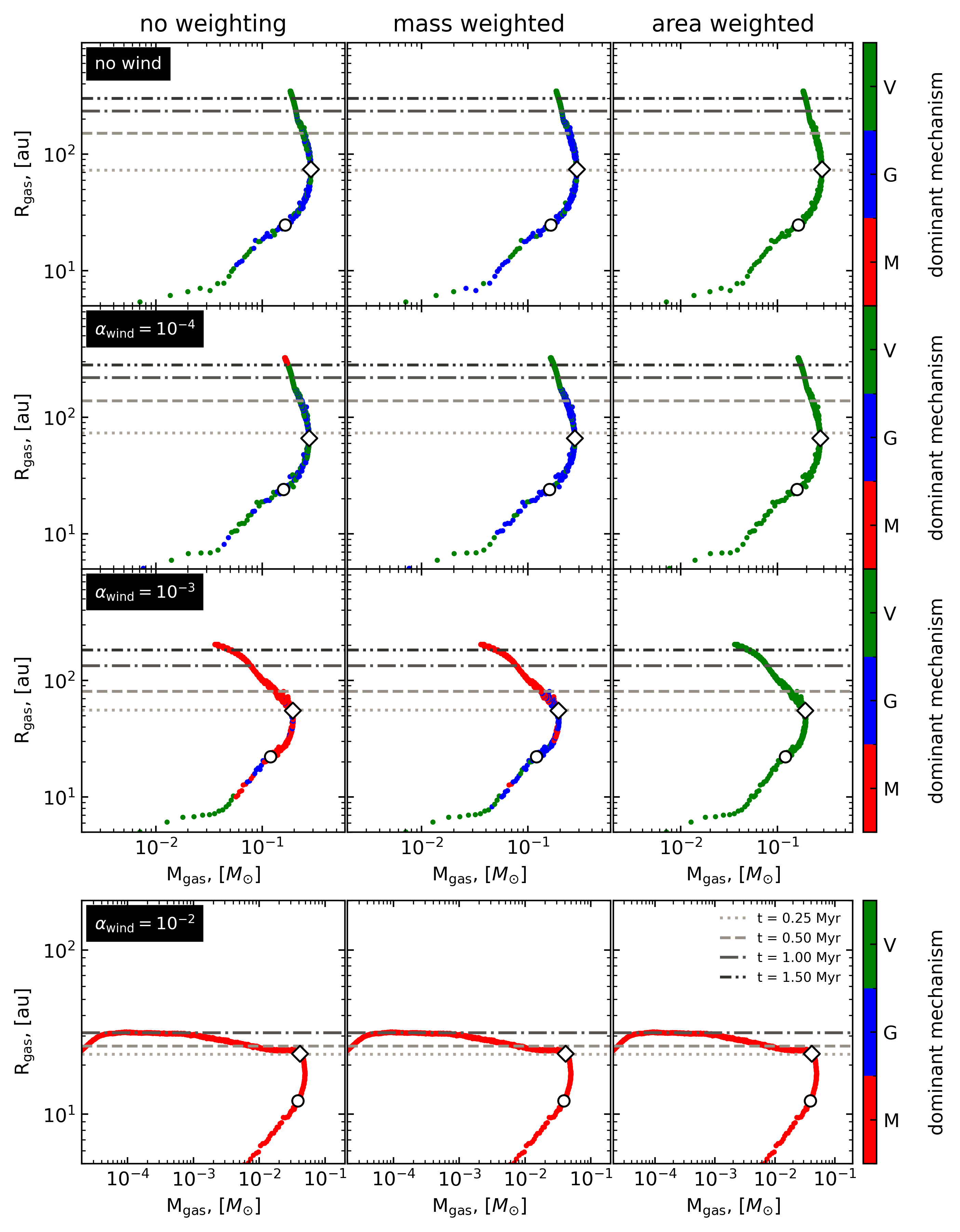}
    \caption{Gas disc radius versus disc mass (up to 2 Myr). The circle marks the time corresponding to the infall of 50\% of the mass from the collapsing envelope, and the diamond marks the time corresponding to 90\% of the mass. The colors indicate the dominant torque at a given time: red -- magnetic, blue -- gravitational, green -- viscous. The dominant mechanism is determined by different methods: left column -- the greatest (by absolute value) net torque in the disc, middle column -- the largest net torque with mass weighting, right column -- the largest net torque with area weighting. The sign is taken into account when summing local torques.}
    \label{fig: gas_RM-diagramm-mod}
\end{figure}

It is interesting to relate the change in disc masses and sizes to the evolution of torques. Fig.~\ref{fig: gas_RM-diagramm-mod} shows the gas disc radius as a function of its mass.
The first intersection of the curve with the dashed, dotted, dash-dotted and dash-dot-dotted lines denotes the time instances $t=$ 0.25 Myr, 0.5 Myr, 1 Myr, and 1.5 Myr, respectively.
The circle and diamond symbols mark the evolutionary stages when the collapsing envelope loses 50\% and 90\% of its mass, respectively, which approximately corresponds to the transition from evolutionary Class 0 to Class~$\rm I$ and then to Class~$\rm II$ \citep{Dunham2010}.
It can be seen that at some point all the plots turn to the left. This happens because the disc mass stops increasing as the infall of material from the envelope ceases at evolutionary Class~$\rm II$; thus, the ``turn'' of the plot generally occurs near the diamond symbol. Subsequently, accretion onto the star and the magnetic disc wind only contribute to the removal of mass from the disc, and the higher the wind efficiency, the faster the mass decreases, and as a result the ``turn'' is more prominent. A decrease in disc size is observed only in the model with $\alpha_{\rm wind} = 10^{-4}$ after 1 million years.

The color of the plot in Fig.~\ref{fig: gas_RM-diagramm-mod} reflects the dominant angular momentum transport mechanism at a given time: red -- magnetic disc wind, blue -- gravity, green -- viscosity.
In the left column (no weighting), the mechanism is considered dominant if the corresponding total disc torque (see Eqs. (\ref{eq: net_magn_torq_disk}) -- (\ref{eq: net_visc_torq_disk})) is greater in absolute magnitude than the others. However, as noted above, such an approach to determining the dominant mechanism can lead to an overestimation of the influence of the magnetic disc wind, since its torque is always negative.
Therefore, approaches to determining the dominant mechanism with mass weighting (middle column, mass weighted) and area weighting (right column, area weighted) were also considered:
\begin{equation}
    \bar{\Gamma}_{\rm mass} = \frac{\sum_{\phi} \sum_{r=3~{\rm au}}^{R_{\rm gas}} \left( \Gamma(r,\phi) \cdot m(r,\phi) \right)}{\sum_{\phi} \sum_{r=3~{\rm au}}^{R_{\rm gas}} m(r,\phi)},
\label{eq: mass_weighted_torque}
\end{equation}
\begin{equation}
    \bar{\Gamma}_{\rm area} = \frac{\sum_{\phi} \sum_{r=3~{\rm au}}^{R_{\rm gas}} \left( \Gamma(r,\phi) \cdot S_{\rm cell}(r,\phi) \right)}{\sum_{\phi} \sum_{r=3~{\rm au}}^{R_{\rm gas}} S_{\rm cell}(r,\phi)},
\label{eq: area_weighted_torque}
\end{equation}
where instead of $\Gamma(r,\phi)$ the corresponding local torques (magnetic, gravitational, viscous) are used.
These approaches take into account not only the magnitude of the torque, but also the fraction of the gas mass or disc area affected by these torques.

From the data presented in the second column of Fig.~\ref{fig: gas_RM-diagramm-mod}, it can be seen that mass weighting slightly enhances the role of gravity, since the local gravitational torque depends on mass (see Eq.~(\ref{eq: grav_torque_loc})).
A comparison of the data presented in the third column of Fig.~\ref{fig: gas_RM-diagramm-mod} with the color map in the bottom panel of Fig.~\ref{fig: disk_torques_colormaps} may give the misleading impression of contradictory results. 
Indeed, from Fig.~\ref{fig: disk_torques_colormaps} it can be seen that during the first 500 kyr of evolution, the gravitational torque locally dominates, even in the outer regions. In this time interval, the net gravitational torque, as can be seen from Fig.~\ref{fig: disk_torques}, also mostly exceeds the viscous torque. However, according to Fig.~\ref{fig: gas_RM-diagramm-mod}, the area-weighted viscous torque turns out to be dominant even at the initial stages of evolution. This can be explained by the fact that in the outer vast regions the viscous torque is predominantly positive, while the gravitational torque is ``noisy'' and often changes sign (see Appendix~\ref{app: torques_colormaps}). Therefore, the area-weighted average value of the gravitational torque turns out to be smaller than that of the viscous torque, although locally (by absolute value) the gravitational torque dominates everywhere. It can be concluded that in the outer regions the action of viscosity is global and unidirectional, ensuring a systematic transport of angular momentum even when the local efficiency of the viscous torque is not the highest.

It can also be seen that in the model with $\alpha_{\rm wind} = 10^{-4}$ the magnetic torque does not have a significant effect on the disc evolution at least until 1.5 Myr; therefore, the differences from the fiducial model (no wind) are minimal. Starting from the model with $\alpha_{\rm wind} = 10^{-3}$, the contribution from the magnetic torque becomes more significant, and in the approach without weighting it begins to dominate even before the material infall from the envelope ceases.
For the model with $\alpha_{\rm wind} = 10^{-2}$, the magnetic wind prevails regardless of the method of determining the dominant mechanism, since in this model its torque is much greater than the other torques at any time.

\subsection{The problem of dust disc depletion}
\label{Subsection: dust_depletion}

The simulation results described in Sections~\ref{Subsection: general_evolution} and \ref{Subsection: masses_and_sizes} indicate that over time a significant depletion of the dust disc occurs, and after 2 million years of evolution in all models, except for $\alpha_{\rm wind} = 10^{-2}$, the dust-to-gas mass ratio drops below $10^{-4}$.
In this section, it is shown how dust depletion can be avoided. To do this, a comparison was made between the original model with $\alpha_{\rm wind} = 10^{-3}$, which uses a fragmentation velocity value of $v_{\rm frag} = 5~{\rm m~s^{-1}}$, and models with a similar set of parameters (see Tab.~\ref{tab: model_parameters}), but lower values of $v_{\rm frag} = 1~{\rm m~s^{-1}}$ and $v_{\rm frag} = 0.5~{\rm m~s^{-1}}$. There is considerable uncertainty in the choice of the threshold value for the fragmentation velocity of dust particles. 
Fragmentation velocities of $v_{\rm frag} \le 1~{\rm m~s^{-1}}$ were obtained in several laboratory experiments \citep{Beitz2011, Fritscher2021} and were previously used
in theoretical modeling \citep{Okuzumi2019}.
However, other experimental and theoretical works indicate that the fragmentation threshold may be higher. 
In particular, \citet{GundlachBlum2015} showed that water-ice particles are capable of sticking at collision velocities up to $\sim 10~{\rm m~s^{-1}}$, and simulations by \citet{Wada2009} demonstrated the growth of ice aggregates at velocities up to $50~{\rm m~s^{-1}}$, though with reducing efficiency.
A value of $v_{\rm frag}=5~{\rm m~s^{-1}}$ was used as a representative parameter in theoretical models of the evolution of protoplanetary discs \citep{Muller2021, Pinilla2025} and is also chosen as a reference value in the present study.

\begin{figure}
    \centering
    \includegraphics[width=\columnwidth]{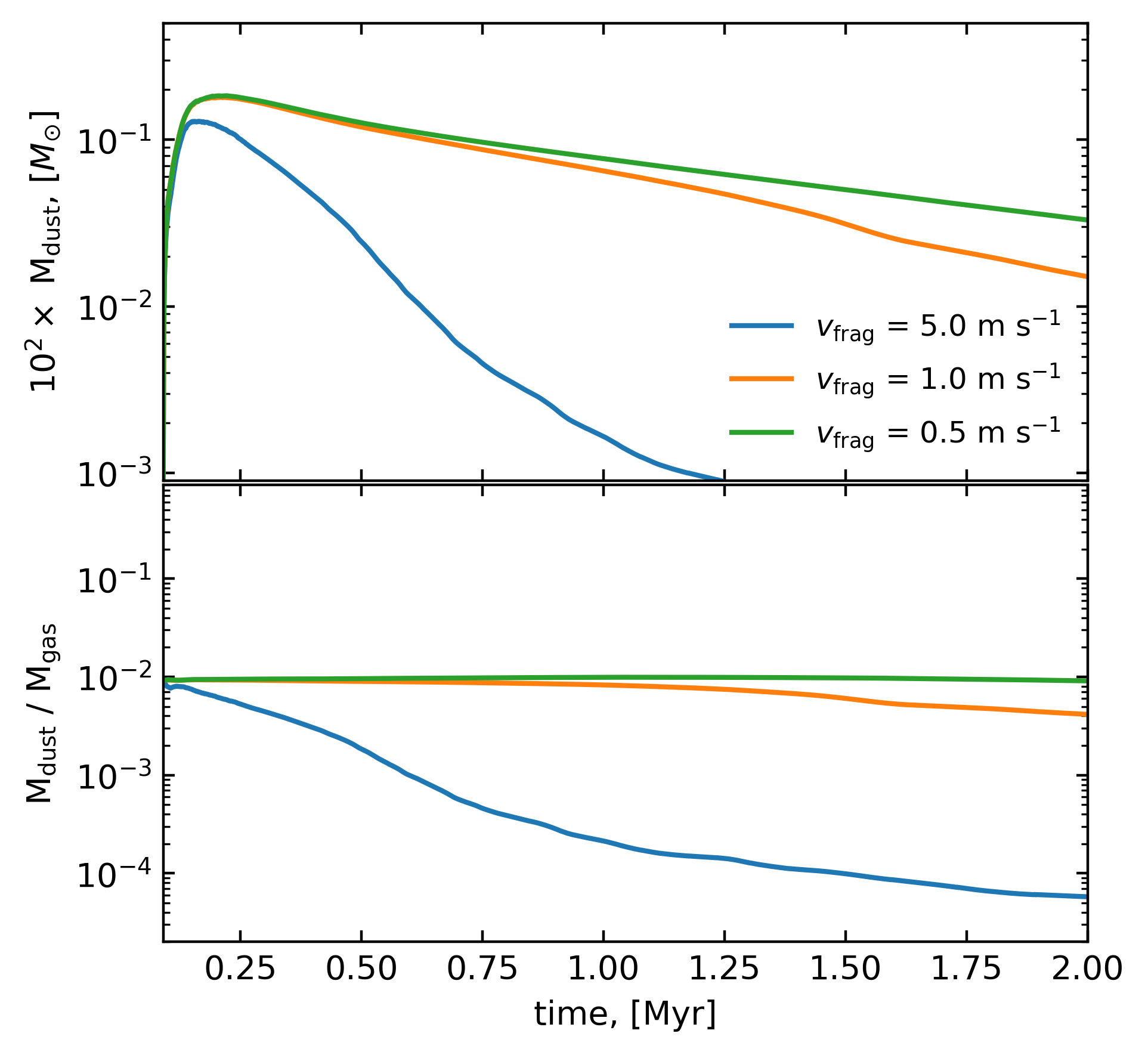}
    \caption{Evolution of the dust disc mass (first panel) and the dust-to-gas mass ratio in the disc (second panel) in models with the wind intensity of $\alpha_{\rm wind} = 10^{-3}$ and different values of the fragmentation velocity $v_{\rm frag}$.} 
    \label{fig: vfrag_dust_masses}
\end{figure}

The first and second panels of Fig.~\ref{fig: vfrag_dust_masses} show the time dependence of the dust mass in the disc and the dust-to-gas mass ratio, respectively. It can be seen that in the models with a lower value of $v_{\rm frag}$ more dust remains in the disc, and the value of $M_{\rm dust} / M_{\rm gas}$ is almost two orders of magnitude higher by the end of simulations. In the model with $v_{\rm frag} = 0.5~{\rm m~s^{-1}}$ it is only slightly below $10^{-2}$ by $t$ = 2 Myr.

\begin{figure*}
    \centering
    \includegraphics[scale=0.6]{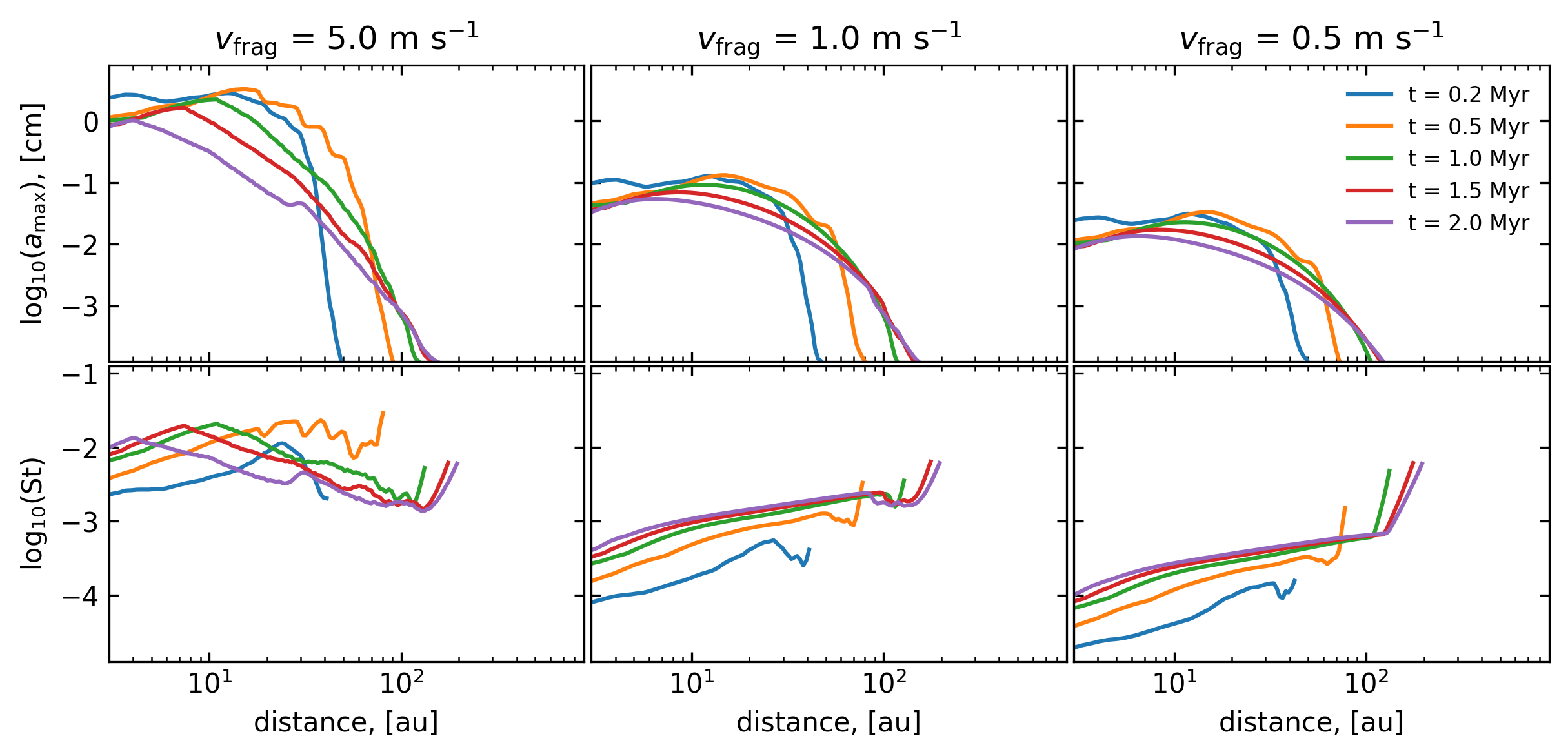}
    \caption{Evolution of the radial profiles of the maximum dust grain size (top panel) and the Stokes number (bottom panel) in models with the wind intensity of $\alpha_{\rm wind} = 10^{-3}$ and different values of the fragmentation velocity $v_{\rm frag}$. Radial profiles of the Stokes number are shown only within the disc radius.}
    \label{fig: vfrag_density_profiles}
\end{figure*}

Since the fragmentation barrier is proportional to the square of the fragmentation velocity $a_{\rm frag} \propto v_{\rm frag}^2$, a decrease in $v_{\rm frag}$ by an order of magnitude leads to a reduction of the maximum dust size by a factor of 100. As can be seen in Fig.~\ref{fig: vfrag_density_profiles} (top panel), in the model with $v_{\rm frag} = 0.5~{\rm m~s^{-1}}$ the dust cannot grow even to 1 mm. Dust grains of smaller size have a lower Stokes number (in the $v_{\rm frag} = 0.5~{\rm m~s^{-1}}$ model, $\mathrm{St}$ does not even reach $10^{-3}$ in the disc), which is confirmed by the bottom panel of Fig.~\ref{fig: vfrag_density_profiles}. Therefore, smaller dust particles are less susceptible to braking by the azimuthal headwind from the gas, so their inward radial drift is significantly suppressed and they are advected mostly with the gas. As a result, more dust remains in the disc although the fraction of dust mass removed by the wind also increases (see Appendix~\ref{app: mass_frac_v_frag}).
Thus, if only small values of $v_{\rm frag} \le 1~{\rm m~s^{-1}}$ are used, catastrophic dust disc depletion might be avoided, which is consistent with the results of other studies, e.g., \citet{Birnstiel2009, Pinilla2025}. 
However, we note that, due to limited numerical resolution, which results in smoothing of the pressure gradients in the spiral structure, our simulations may underestimate the efficiency of dust retention by spiral arms (see Appendix~\ref{app: char_drift_timescales}), and, as a result, overestimate dust depletion in the case of $v_{\rm frag} = 5~{\rm m~s^{-1}}$. Besides, note that we do not consider other dust retention mechanisms, such as viscosity variations in the disc (e.g., ``dead zones''), pressure bumps near snow lines, vertical shear instability, tidal interactions with stellar companions and flybys, planet-disc interaction (see \citet{Bae2023} for a broader review of dust trapping).

\section{DISCUSSION}
\label{Section: discussion}

In this study, we analysed the influence of various angular momentum transport mechanisms on the evolution of a protoplanetary disc. However, we note that the effects of these mechanisms  are represented by a combination of direct self-consistent calculations (gravity) and parameterized models (turbulent viscosity and magnetic disc wind). Below, we outline the limitations of our approach and carry out a comparison with previously published works.

{\it Limitations of the model.} The magnetic disc wind model used in this work follows the approach of \citet{Tabone2022} and includes two free parameters: the magnetic lever arm $\lambda$ and the parameter $\alpha_{\rm wind}$, which characterizes the wind efficiency. The choice of $\lambda = 2$ is based on the interval of characteristic values of this quantity ($1.6$--$5$) given in \citet{Tabone2022}. We set $\alpha_{\rm wind}$ equal to $10^{-2}$, $10^{-3}$, and $10^{-4}$, depending on the model, so that its values are distributed symmetrically around the viscous $\alpha_{\rm visc}=10^{-3}$.
One limitation of our model is that these parameters are not variable in space and time. More sophisticated models \citep[e.g.][]{Lesur2021-1D} treat the properties of the wind as functions of the disc magnetisation, rather than being prescribed through constant parameters, and future developments of our model will follow a similar path.

We note that although $\alpha_{\rm wind}$ is constant, the effects of the magnetic disc wind as represented  by the magnetic torque and the mass-loss rate due to the wind, change in strength over time and space. As can be seen from equations (\ref{eq: ang_mom_loss_wind})--(\ref{eq: sigma-dot-wind}), both $T^{\rm w}_{z\phi}$ and $\dot{\Sigma}_{\rm w}$ depend on local characteristics in the disc, such as surface density and temperature, which also vary over time and space.

Variations in the values of $\alpha_{\rm wind}$ can be expected if the underlying magnetic field also varies over time and space. In future studies, we plan to introduce a variable value by analogy with that proposed in \citet{Kadam2025} (see their eqs.~25 and 26).
However, even in our simplified case, 
an approximately constant value of $\alpha_{\rm wind}$ can be retrieved if the magnetic field declines with radius slower than the gas surface density since $\alpha_{\rm wind} \propto B_z^2 / \Sigma_{\rm g}$ \citep{Suzuki2016}. 

For the kinematic viscosity (see Eq.~\ref{eq: viscosity}), the classical $\alpha$-parametrization \citep{ShakuraSunyaev1973} is used. In all simulations, an intermediate value of $\alpha_{\rm visc} = 10^{-3}$ is chosen \citep{Rafikov2017, Rosotti2023}, which is the same at all radial distances and does not change with time. We note, however, that the parameter $\alpha_{\rm visc}$ may not be constant, especially when non-ideal MHD effects are taken into account \citep{Flock2015, Delage2022}. This may lead to the formation of regions in the disc with suppressed viscosity, the so-called ``dead zones'' \citep{Gammie1996}, not considered in our model.

{\it Comparison with previously published works.}
We begin by comparing our two-dimensional approach to simulating the disc wind, which is based on the model of \citet{Tabone2022},  with published one-dimensional models of \citet{Suzuki2016} \citep[see also][]{Kunitomo2020, Weder2023} and \citet{Chambers2019} \citep[see also][]{Alessi&Pudritz2022}. However, it should be noted that none of these works accounts for the self-gravity of the disc. In contrast, our simulations include a self-consistent calculation of the disc's gravitational potential, which allows us to directly account for the influence of gravitational torques on the evolution of the protoplanetary disc. The contribution of gravity is expected to be significant at the initial stages of evolution for as long as matter continues to infall from the collapsing envelope \citep{Vorobyov2009}.

The model of \citet{Suzuki2016} employs a parametrization of turbulent viscosity and magnetic disc wind via a viscosity parameter $\overline{\alpha_{r\phi}}$, a disc wind torque parameter $\overline{\alpha_{z \phi}}$, and a disc wind mass flux parameter $C_{{\rm w},0}$. We emphasize that their definition of the parameter $\overline{\alpha_{z \phi}}$ does not coincide with our $\alpha_{\rm wind}$ (see Appendix~\ref{app: relations_between_wind_models}). The authors consider both cases where $\overline{\alpha_{z \phi}}$ remains constant (or equal to zero), and cases where this parameter depends on the gas surface density. The mass-loss rate due to the wind is characterized by the quantity $C_{\rm w} = \min(C_{{\rm w},0}, C_{{\rm w},{\rm e}})$, where $C_{{\rm w},{\rm e}}$ is calculated from energy consideration and, therefore, may also vary in space and time. \citet{Suzuki2016} conclude that, depending on the specific set of parameters, the evolution of the surface density profiles over time can differ significantly, which is also the case in our models with different $\alpha_{\rm wind}$.
In particular, in their ``Strong DW + constant torque'' case, which is closest to our model with $\alpha_{\rm wind} = 10^{-3}$
(in this case $\overline{\alpha_{z \phi}}$ is fixed and $C_{\rm w}$ is approximately constant beyond 1~au), a similar behavior of the surface density profile is observed (see Fig.~\ref{fig: density_profiles} in our work and Fig.~5 in theirs). 
A direct comparison with the other cases described in \citet{Suzuki2016} is difficult to conduct due to the complex spatial and temporal dependence of the mass-loss parameter $C_{\rm w}$. We note, however, that  $\overline{\alpha_{z \phi}}$ is almost radially constant beyond $10$~au over the $1$~Myr time interval and corresponds to our $\alpha_{\rm wind} = 10^{-4}$.

The model of \citet{Chambers2019} uses three fixed parameters: $v_0$ -- the inward velocity at $1$~au, $f_{\rm w}$ -- the fraction of the velocity $v_0$ caused by the magnetic disc wind, and $K$ -- the efficiency of the mass loss due to the wind.
These parameters can be related to our $\alpha_{\rm visc}$ and $\alpha_{\rm wind}$ (see Appendix~\ref{app: relations_between_wind_models}). A comparison of the ``Purely Viscous Disc'' case with $v_0 = 30$~cm~s$^{-1}$ and $f_{\rm w} = 0$, corresponding to ours $\alpha_{\rm visc} \simeq 10^{-2}$ and $\alpha_{\rm wind} = 0$, and the ``Wind Dominated Disc with a Slow Wind'' case with $v_0 = 30$~cm~s$^{-1}$, $f_{\rm w} = 0.8$, and $K = 1$, corresponding to ours $\alpha_{\rm visc} \simeq 2 \times 10^{-3}$ and $\alpha_{\rm wind} \simeq 2.5 \times 10^{-3}$, allowed \citet{Chambers2019} to conclude that viscous spreading is reduced in the presence of a wind, which is consistent with our results. We also found a qualitatively and quantitatively similar behavior of the gas surface density profiles in the ``Wind Dominated Disc with a Slow Wind'' case and our model with $\alpha_{\rm wind} = 10^{-3}$.
Comparison with other models of \citet{Chambers2019} is complicated due to the use of nonidentical initial conditions and viscosity that varies from model to model.

A comparison with simplified models that avoid using parametrization is possible only on a qualitative level. Among such works, \citet{Bai2016} can be mentioned, where the distribution and evolution of the magnetic flux threading the disc are considered, and the influence of the magnetic disc wind is calculated from physical conditions. Viscosity is also included in this model; however, in the bulk of the disc it is very small and does not significantly affect the disc evolution. Qualitatively, this case can be related to our model with $\alpha_{\rm wind} = 10^{-2}$, in which the magnetic torque is absolutely dominant. \citet{Bai2016} emphasizes, in addition to the suppression of viscous spreading, that the disc loses roughly the same amount of mass through the wind as through accretion onto the star. For our model with $\alpha_{\rm wind} = 10^{-2}$ the same outcome can be concluded, which is confirmed by the accretion rate onto the star being almost identical to the total disc mass-loss rate due to the wind (see Appendix~\ref{app: accretion_rates} and Fig.~\ref{fig: accr_rates}). Moreover, the substantial increase in the dust-to-gas mass ratio reported by \citet{Bai2016} is observed in our results with the strongest wind as well (see Fig.~\ref{fig: gas_and_dust_radii_masses}).

In a recent study of the effects of magnetic disc wind,  \citet{Kadam2025} also simulate viscosity and gravity in a way similar to that in the present paper, but they use a more detailed model of the magnetic disc wind. The approximation formulas for the wind quantities are based on the results of local shearing-box simulations of \citet{Bai2013-2}. The mass-loss rate and the magnetic torque in this model depend not only on the local conditions in the disc, but also on the properties of the star and the evolution of the magnetic field in the flux-freezing approximation. Although \citet{Kadam2025} do not provide a direct quantitative comparison of the contributions of the various torques, their main conclusion that the inclusion of a magnetic wind reduces the disc mass and size on $\sim 1$~Myr timescales is in agreement with our results. 

A direct comparison of our model with the results of 3D simulations is difficult, since the latter usually cover a relatively short period of disc evolution ($\sim 10^4$--$10^5$~yr). Nevertheless, some general patterns can be identified in the initial stages after disc formation. \citet{Xu2021} showed that during the first few kyr of disc evolution, gravitational instability is the main mechanism of angular momentum transport, whereas the contribution of the magnetic disc wind is negligible at this stage. A similar picture is observed in all our models, with the exception of the case with $\alpha_{\rm wind} = 10^{-2}$, where the magnetic torque exerts a non-negligible influence even at the earliest stages. Numerical simulations of \citet{Machida2024}, covering the evolution up to $t \sim 1.5 \times 10^5$~yr, confirm that as the collapsing envelope depletes and the gravitational torque weakens, the magnetic disc wind becomes dominant. The overall evolutionary picture -- a transition from the prevalence of gravitational torques to that of magnetic ones -- is consistent with our conclusions, but our models can in addition identify the disc regions with the prevalence of turbulent viscosity.

\section{CONCLUSIONS}
\label{Section: conclusions}

In this work, based on global numerical hydrodynamic simulations in the thin-disc limit, the evolution of a protoplanetary disc under the influence of three angular momentum transport mechanisms was investigated: gravitational instability, turbulent viscosity, and magnetic disc wind.
The simulations were carried out over timescales comparable to the characteristic disc lifetime (up to 2 Myr), making it possible to track the disc evolution from the embedded phase to subsequent stages. 
The approach used, in contrast to one-dimensional semi-analytical and numerical models, allowed the self-consistent calculation of gravitational torques and a comparison of the contribution of each mechanism to mass and angular momentum transport within a single framework. The main results of the work can be formulated as follows.
\begin{enumerate}
\item All three mechanisms -- gravity, viscosity, and magnetic disc wind -- play an important role in the evolution of a protoplanetary disc: their contributions vary significantly in different disc regions and at different stages of the disc's evolution. Gravitational torques dominate almost throughout the entire disc at the earliest stages, but over time the wind and viscosity become increasingly important (the latter especially in the outer regions). The results obtained confirm that none of the mechanisms can be considered mutually exclusive, and their combined consideration is required for a realistic description of disc evolution.
\item It is shown that, regardless of the wind parameters, a noticeable decrease in the disc mass begins only after the termination of the embedded phase, when gravitational instability ceases to be the dominant mechanism of mass and angular momentum transport; however, the higher the magnetic wind intensity, the faster the disc mass decreases. In the model with $\alpha_{\rm wind} = 10^{-2}$, disc depletion occurs too quickly, at $t \approx 1.25$ Myr, which is inconsistent with the typical lifespan of protoplanetary discs \citep{Mamajek2009, Pfalzner2026}.
\item It is confirmed that in the presence of non-negligible viscosity ($\alpha_{\rm visc} \sim 10^{-3}$), only the most intense magnetic disc wind ($\alpha_{\rm wind} \sim 10^{-2}$) is capable of suppressing the viscous spreading of the disc. In models with a weaker wind, viscosity continues to dominate at late times and results in the disc radius increasing over time, as expected from viscous evolution.
\item The dust-poor disc wind does not prevent the dust-to-gas mass ratio from dropping significantly below the initial value of $10^{-2}$, unless the wind strength is very high. Despite the fact that the spiral structure of the disc contributes to the temporary retention of dust, low values of the fragmentation velocity, $v_{\rm frag} \le 1.0~{\rm m~s^{-1}}$, are required to prevent the rapid exhaustion of dust without additional mechanisms of dust trapping.
\end{enumerate}

Future work will be focused on investigating the evolution of volatiles, and in particular, on the evolution of the C/O ratio, in discs with magnetic wind.
Finally, an important step in future research will be a direct comparison of the disc masses and sizes predicted by our model with observational data.

\section*{ACKNOWLEDGEMENTS}

We are thankful to the anonymous referee for careful reading, constructive comments and suggestions that helped to improve the manuscript.
E.R. and E.V. acknowledge support by the Ministry of Science and Higher Education of the Russian Federation (State contract FENW-2026-0028).
P.P. and F.G. acknowledge funding from the UK Research and Innovation (UKRI) under the UK government’s Horizon Europe funding guarantee from ERC (under grant agreement No 101076489).
Simulations were performed on the Austrian Scientific Cluster (\href{https://asc.ac.at/}{https://asc.ac.at/}).

\section*{Data Availability}
The datasets generated and analysed during the current study are available from the leading author upon request.

\bibliographystyle{mnras.bst}
\bibliography{main.bib}
\clearpage

\appendix

\section{Comparison of disc evolution with different methods for calculating the gravitational potential}
\label{app: grav_pot_calc}

\begin{figure}
    \centering
    \includegraphics[width=\columnwidth]{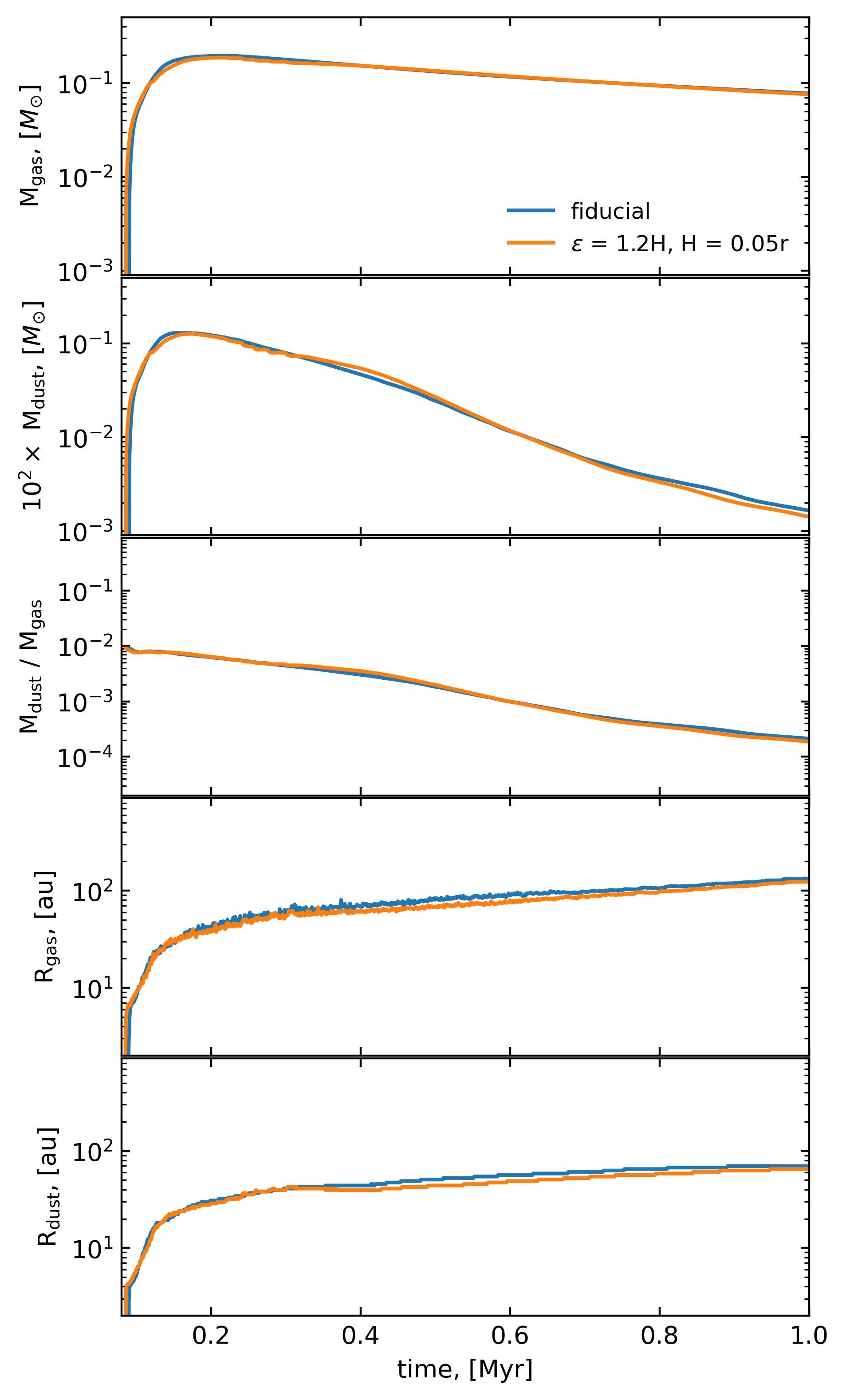}
    \caption{Evolution (from top to bottom) of the gas disc mass, dust disc mass, dust-to-gas mass ratio, gas disc radius, and dust disc radius in the model without a smoothing parameter when calculating the gravitational potential (fiducial) and in the model using the $\epsilon$-parameter according to formula (\ref{eq: grav_potential_with_eps_app}).}
    \label{fig: GRAV_masses_evolution}
\end{figure}

Fig.~\ref{fig: GRAV_masses_evolution} shows the time evolution (up to 1 Myr) of the masses and radii of the gas and dust discs, as well as the dust-to-gas mass ratio, for two models. The first model (fiducial) is the model used in this work with a wind intensity of $\alpha_{\rm wind} = 10^{-3}$, where the gravitational potential is calculated without introducing an $\epsilon$-smoothing parameter, as described in \citet{Vorobyov2024}, based on \citet{Binney1987}. The second model has an identical set of parameters, except for the approach to calculating the gravitational potential, which is computed with smoothing. Instead of the standard expression (\ref{eq: grav_potential}) for the gravitational potential, this approach uses its modified version, which makes it possible to avoid the singularity at $r=r'$ and $\phi=\phi'$, when the denominator becomes equal to zero and the integral diverges. The expression for the gravitational potential with the $\epsilon$-smoothing parameter is written as
\begin{equation}
    \Phi(r, \phi) = -G \int^{r_{\rm out}}_{r_{\rm in}} r' dr' \int^{2 \pi}_{0} \frac{\left(\Sigma_{\rm g}(r', \phi') + \Sigma_{\rm d,tot}(r', \phi') \right) d\phi'}{\sqrt{(r')^2 + r^2 - 2 r r' \cos(\phi'-\phi) + \epsilon^2}}.
\label{eq: grav_potential_with_eps_app}
\end{equation}
To avoid mathematical difficulties when using the convolution method, it is recommended to use a specific form of the $\epsilon$-parameter dependence, proportional to the radial distance in the disc, i.e., $\epsilon \propto r$ \citep{Baruteau2008}. Usually, the $\epsilon$-parameter is related to the disc vertical scale height $H$, since in a protoplanetary disc it is proportional to the radial distance as well. The choice of parameters $\epsilon = 1.2H$ and $H = 0.05r$ for the second model is justified in \citet{Muller2012}.

As can be seen in Fig.~\ref{fig: GRAV_masses_evolution}, no significant differences are observed in the global disc evolution for the models considered. This confirms the validity of the method for calculating the gravitational potential used in this study, which does not violate Newton's law of gravity caused by the introduction of the $\epsilon$-parameter.

\section{Comparison of accretion rates onto the star and rates of matter removal by the magnetic disc wind}
\label{app: accretion_rates}

To launch a magnetic disc wind, energy is required, which is drawn from the gravitational energy of the accretion flow of matter moving through the disc towards the star.

\begin{figure}
    \centering
    \includegraphics[width=\columnwidth]{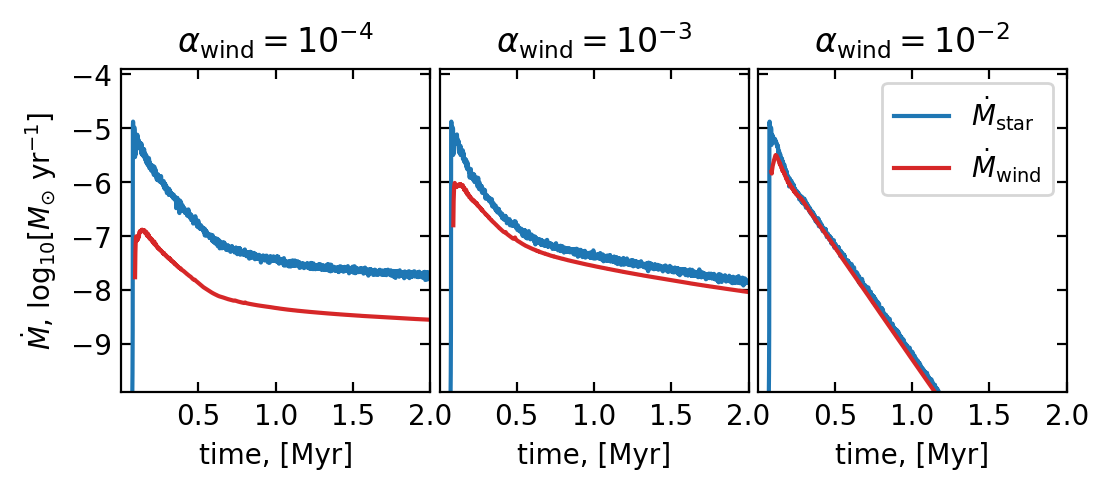}
    \caption{Time evolution of the accretion rate onto the star (blue curve) and the total mass loss rate in the disc due to the wind (red curve).} 
    \label{fig: accr_rates}
\end{figure}

Fig.~\ref{fig: accr_rates} shows the time dependence of the accretion rate onto the star $\dot{M}_{\rm star}$, which in the quasi-stationary approximation can be regarded as the rate of matter transport through the disc, and the rate of disc mass loss due to the magnetic disc wind $\dot{M}_{\rm wind}$. The latter quantity is calculated as the sum of the local rates of mass removal by the wind:
\begin{equation}
   \dot{M}_{\rm wind} = \sum_{\phi} \sum_{r} \dot{\Sigma}_{\rm wind}(r,\phi) S_{\rm cell}(r, \phi),
\label{eq: dotM_wind}
\end{equation}
where $S_{\rm cell}$ is the cell area. It can be seen that in all models used in this work even in the model with the most intense wind, $\alpha_{\rm wind} = 10^{-2}$, the total mass loss rate due to the wind (red curve) does not exceed the accretion rate onto the star (blue line). Thus, the energy principle, namely that the wind is powered by the energy of the accretion flow through the disc (accretion-powered wind), is not violated in the wind model used.

\section{Maximum dust size evolution}
\label{app: amax_evolution}

The fragmentation barrier $a_{\rm frag}$ and the radial drift barrier $a_{\rm drift}$ limit the maximum size to which dust can grow in a protoplanetary disc. These quantities are defined as follows:
\begin{equation}
    a_{\rm frag} = \frac{2 \Sigma_{\rm g} v_{\rm frag}^2}{3 \pi \rho_{\rm s} \alpha_{\rm visc} c_{\rm s}^2},
\label{eq: a_frag}
\end{equation}
\begin{equation}
    a_{\rm drift} = \frac{2 \Sigma_{\rm d,gr} r^2}{\pi \rho_{\rm s} H_{\rm g}^2} \gamma^{-1},
\label{eq: a_drift}
\end{equation}
where $v_{\rm frag}$ is the fragmentation velocity, $\rho_{\rm s}$ is the material density of dust ($\rho_{\rm s} = 2.24\ {\rm g\ cm^{-3}}$), and the magnitude of the logarithmic pressure gradient $\gamma$ is calculated using:
\begin{equation}
    \gamma = \left| \frac{d\ {\rm ln} \mathcal{P}}{d\ {\rm ln} r} \right|.
\label{eq: pressure_rad_slope}
\end{equation}

\begin{figure}
    \centering
    \includegraphics[width=\columnwidth]{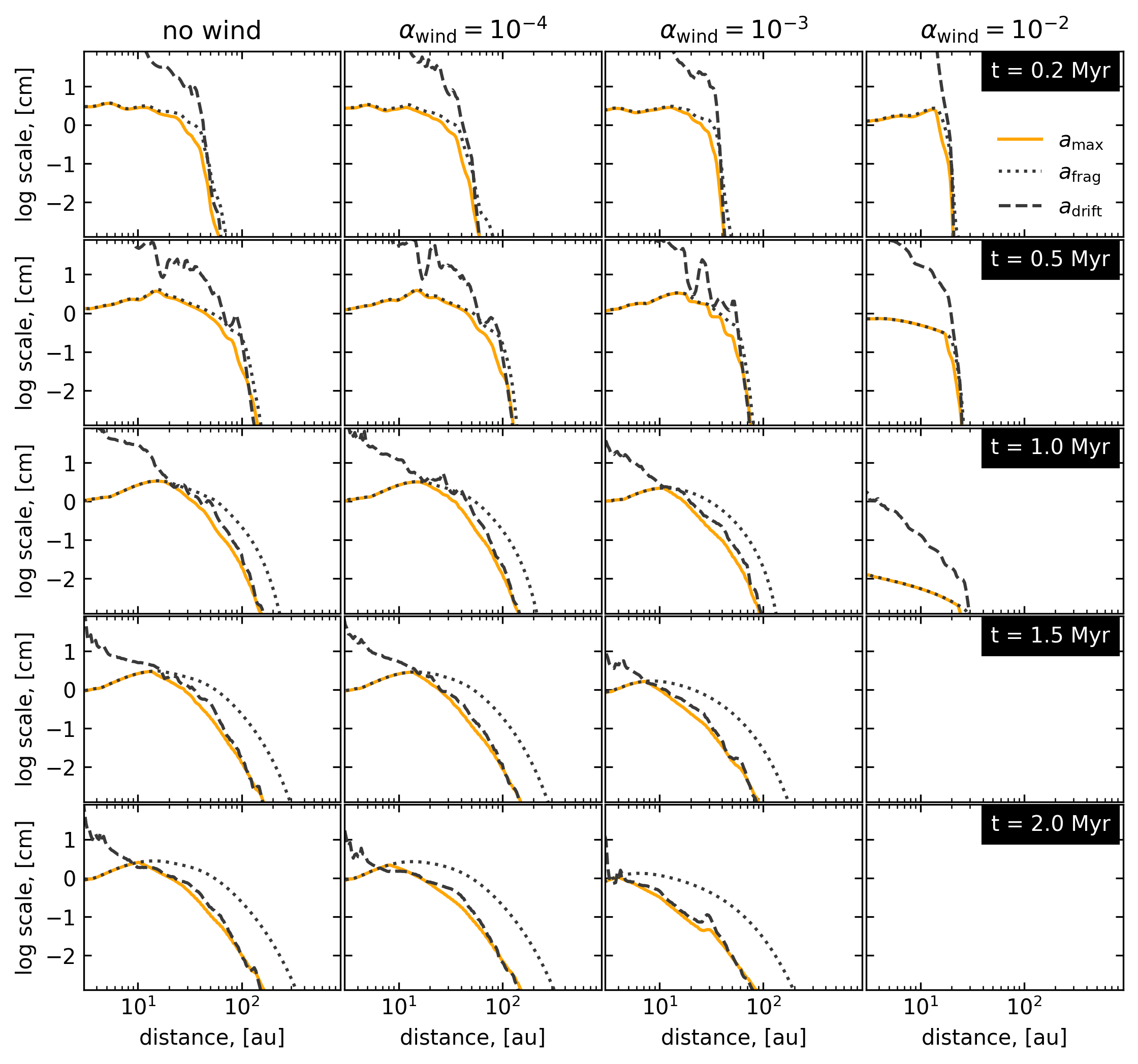}
    \caption{Radial profiles of the azimuthally averaged maximum dust size (solid orange line) at different times. The fragmentation barrier is shown by the dashed line, the radial drift barrier -- by the dotted line.}
    \label{fig: amax_profiles}
\end{figure}

Fig.~\ref{fig: amax_profiles} shows the evolution of the radial profiles of the azimuthally averaged maximum dust size $a_{\rm max}$, and also plots the profiles of $a_{\rm frag}$ (dashed line) and $a_{\rm drift}$ (dotted line). 
It is noticeable that the maximum dust size, calculated with our model, is in good agreement with the analytical estimates of the dust growth and drift barriers.
In all models, except for the model with $\alpha_{\rm wind} = 10^{-2}$, up to $t = 0.5$ Myr, the dust size is limited predominantly by the fragmentation barrier; however, over time the radial drift barrier begins to play an increasingly significant role, and by $t = 2$ Myr it is precisely $a_{\rm drift}$ that restricts dust growth in most of the disc. Equations (\ref{eq: a_frag}) and (\ref{eq: a_drift}) show that the fragmentation barrier is proportional to the gas density, while the radial drift barrier depends on the density of grown dust. Since the dust-to-gas ratio $\xi_{\rm d2g}$ decreases over time (see Fig.~\ref{fig: density_profiles}), $a_{\rm drift}$ becomes smaller than $a_{\rm frag}$. The lower values of $a_{\rm drift}$ compared to the fiducial model explain the decrease in the maximum dust size in the models with higher wind intensity. A quantitative analysis shows that $\Sigma_{\rm d,gr}$ has the greatest influence on the magnitude of the radial drift barrier. Since at the same radial distance in the models with a wind the grown dust density is lower, the maximum dust grain size reaches the radial drift barrier at smaller values.

\section{Characteristic drift timescale of grown dust}
\label{app: char_drift_timescales}

Since by our assumption the wind does not carry away the mass-dominant grown dust, the dust disc depletion can be explained by the drift of grown dust relative to the gas with subsequent accretion onto the star. To verify this, we analyse the characteristic radial drift timescales of grown dust, which at a radial distance $r$ are calculated as
\begin{equation}
   \tau_{\rm drift} = \frac{r}{|u_r - v_r|},
\label{eq: tau_drift}
\end{equation}
where $u_r$ and $v_r$ are the radial components of the dust and gas velocities, respectively, calculated during the simulations.
\begin{figure*}
    \centering
    \includegraphics[scale=0.5]{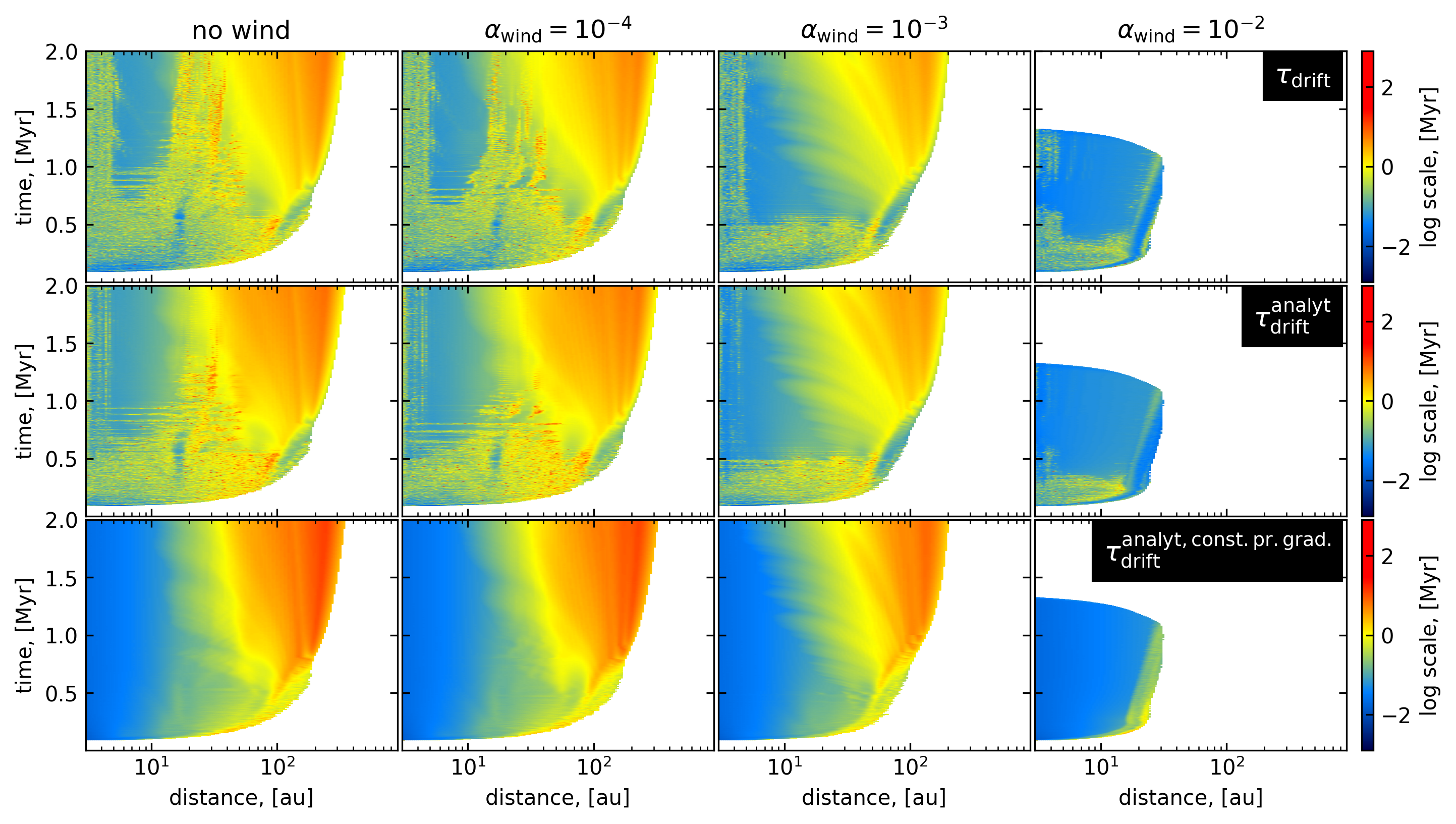}
    \caption{Evolution of the azimuthally averaged characteristic radial drift timescales of grown dust $\tau_{\rm drift}$ calculated using different radial dust velocities: from the simulations (top panel), analytically as the sum of the advective and gradient components from the model data (middle panel), and analytically as the sum of the advective and gradient components but with a constant pressure gradient (bottom panel).} 
    \label{fig: drift_timescales}
\end{figure*}

As can be seen from the top panel of Fig.~\ref{fig: drift_timescales}, which shows the temporal evolution of the characteristic radial drift time calculated from radial gas and dust velocities obtained from simulations, in most of the disc the values of $\tau_{\rm drift}$ lie within the range of $10^{4} - 10^{6}~{\rm yr}$; therefore, over the considered disc evolution time (2 Myr) the dust has sufficient time to separate from the gas and accrete onto the star, which leads to a decrease in $\xi_{\rm d2g}$ in all models except for the model with $\alpha_{\rm wind} = 10^{-2}$. In the latter, the decrease in the dust-to-gas ratio is less prominent, since with characteristic dust drift times comparable to the other models, the intense decrease in the gas surface density in this model compensates for the reduction in the dust-to-gas ratio caused by dust drift.

We note that the characteristic radial drift time of grown dust calculated from the model gas and dust velocities (top panel of Fig.~\ref{fig: drift_timescales}) is in good agreement with the analytical estimate of the characteristic radial drift time (middle panel of Fig.~\ref{fig: drift_timescales}).
The radial drift velocity of grown dust can be divided into two components: the gradient drift velocity $u_{r, {\rm grad}}$ and the advective drift velocity $u_{r, {\rm adv}}$. The latter can be calculated using
\begin{equation}
    u_{r, {\rm adv}} = \frac{v_r}{1 + {\rm St}^2},
\label{eq: u_adv}
\end{equation}
therefore, at low Stokes numbers ${\rm St} \ll 1$, the advective drift velocity does not lead to separation of gas and dust. To estimate the gradient drift velocity, the following analytical approximation can be used:
\begin{equation}
    u_{r, {\rm grad}} = - \frac{2 V_{\rm K} {\rm St} \eta_{\rm dev}}{1 + {\rm St}^2},
\label{eq: u_grad}
\end{equation}
where $V_{\rm K}$ is the Keplerian velocity, and $\eta_{\rm dev}$ characterizes the deviation of the gas disc rotation from Keplerian; it is proportional to the pressure gradient in the disc and is defined as follows:
\begin{equation}
   \eta_{\rm dev} = - \frac{1}{2} \left( \frac{H_{\rm g}}{r} \right)^2 \frac{d\ {\rm ln} \mathcal{P}}{d\ {\rm ln} r}.
\label{eq: eta_deviation}
\end{equation}
Then the characteristic radial drift time $\tau_{\rm drift}^{\rm analyt}$ is defined similarly to formula (\ref{eq: tau_drift}):
\begin{equation}
   \tau_{\rm drift}^{\rm analyt} = \frac{r}{|u_{r, {\rm analyt}} - v_r|},
\label{eq: tau_drift_analyt}
\end{equation}
where instead of the dust velocity $u_r$ calculated self-consistently during the simulations, the velocity $u_{r, {\rm analyt}} = u_{r, {\rm adv}} + u_{r, {\rm grad}}$ is used. When calculating the velocity components using formulas (\ref{eq: u_adv}) and (\ref{eq: u_grad}), the simulation data are used.

However, the characteristic radial drift times in the top and middle panels of Fig.~\ref{fig: drift_timescales} differ significantly from the characteristic radial drift time presented in the bottom panel. In calculating the latter, the same formulas (\ref{eq: u_adv})--(\ref{eq: tau_drift_analyt}) were used as for $\tau_{\rm drift}^{\rm analyt}$, but a constant pressure gradient ($d\ {\rm ln} \mathcal{P} / d\ {\rm ln} r = -2.75$) for a standard protoplanetary disc \citep{Birnstiel2024}, was adopted in formula (\ref{eq: eta_deviation}). The difference between the characteristic times is especially noticeable during the period when the infall of material from the collapsing envelope onto the protoplanetary disc continues and gravitational instability plays a major role, i.e., during the first half a million years of evolution. The characteristic drift time from the simulation results can be greater than the analytical estimate with a constant pressure gradient due to the influence of the spiral structure, which can contribute to the temporary retention of dust \citep{Rice2025}. However, for a detailed study of this effect, it is necessary to carry out simulations with higher numerical resolution, which will be performed in future studies.

\section{Evolution of dust disc radius limited by dust surface density}
\label{app: dust_disc_radii}

\begin{figure}
    \centering
    \includegraphics[width=\columnwidth]{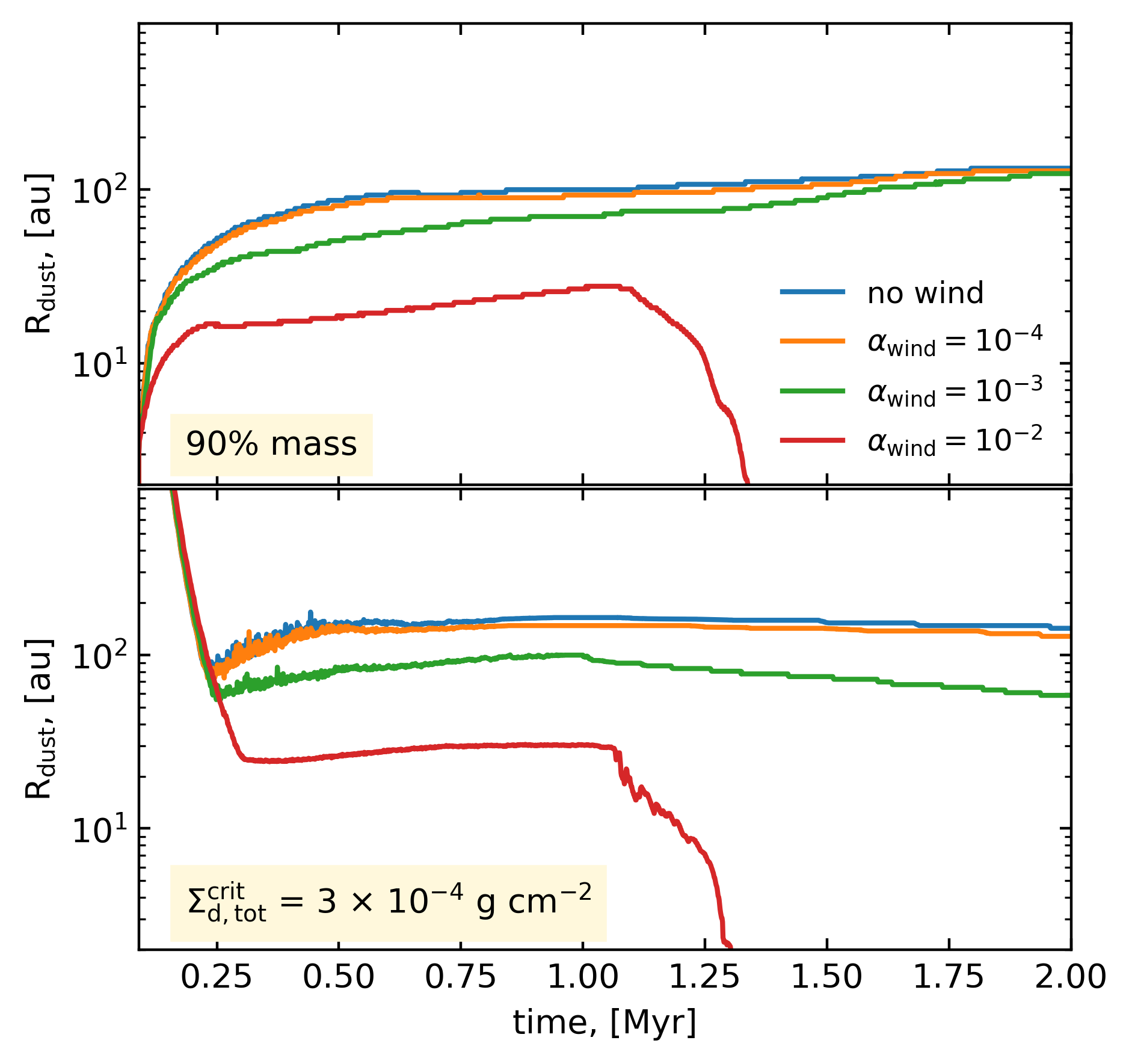}
    \caption{Evolution of the dust disc radii. Top panel: $R_{\rm dust}$ determined as a boundary of the region where 90\% of the total dust mass resides (as described in Section~\ref{Subsection: masses_and_sizes}). Bottom panel: $R_{\rm dust}$ is limited by a critical value of the total dust surface density $\Sigma_{\rm d, tot}^{\rm crit} = 3 \times 10^{-4}~{\rm g~cm^{-2}}$.}
    \label{fig: dust_disc_rad_crit_v}
\end{figure}

In Section~\ref{Subsection: masses_and_sizes} we demonstrated that the dust disc radii $R_{\rm dust}$ (determined as a boundary of the region where 90\% of the total dust mass resides) tend to expand over time, except for the model with the most intense wind ($\alpha_{\rm wind} = 10^{-2}$). This trend, however, may change if other definitions of $R_{\rm dust}$ are applied. For instance, if the radius is limited by a critical value of the total dust surface density $\Sigma_{\rm d, tot}^{\rm crit} = 3 \times 10^{-4}~{\rm g~cm^{-2}}$, which is a factor of 100 lower than the critical value $\Sigma_{\rm crit}$ chosen in our model to limit the \textit{gas} disc radius, then the growth in $R_{\rm dust}$ is saturated around 1.0~Myr and the dust disc radius begins to decline afterwords, see the bottom panel of Fig.~\ref{fig: dust_disc_rad_crit_v}. 
This behavior of $R_{\rm dust}$ can be explained by a decrease in the total dust surface density profile over time due to inward dust drift, somewhat offset by a simultaneous flattening of its slope due to viscous disc spreading.

\section{Mass transport by viscosity}
\label{app: viscous_transport}

The local torque per unit disc surface area created by turbulent viscosity, following the definition
\begin{equation}
    \Gamma_{\rm visc}(r, \phi) = r (\nabla \cdot {\bf{\Pi}})_{\phi},
\label{eq: visc_torque_loc_app}
\end{equation}
is equal to zero if the $\phi$-component of the divergence of the viscous stress tensor
\begin{equation}
    (\nabla \cdot {\bf{\Pi}})_{\phi} = \frac{\partial}{\partial r} \Pi_{\phi r} + \frac{1}{r} \frac{\partial}{\partial \phi} \Pi_{\phi \phi} + \frac{2}{r} \Pi_{r \phi},
\label{eq: nabla_Pi_phi_app}
\end{equation}
vanishes as well. Neglecting the derivatives with respect to the $\phi$-coordinate, for the axisymmetric case the only component affecting the magnitude of the viscous torque is $\Pi_{r\phi}$, given by
\begin{equation}
    \Pi_{r\phi} = \mu r \frac{d \Omega}{d r},
\label{eq: Pi_rphi}
\end{equation}
where $\mu$ is the dynamic viscosity, and $\Omega$ is the angular velocity in the disc. Substituting equation (\ref{eq: Pi_rphi}) into formula (\ref{eq: nabla_Pi_phi_app}) and equating the result to zero, we obtain:
\begin{equation}
    \frac{d}{dr} \left( \mu r \frac{d \Omega}{dr} \right) + 2 \mu \frac{d \Omega}{dr}= 0.
\label{eq: dyn_visc_eq_app}
\end{equation}
Taking into account that in a Keplerian disc $\Omega = \sqrt{G M_{\rm star} / r^{3}}$, this equality reduces to a differential equation for $\mu$:
\begin{equation}
    \frac{d \mu}{dr} = - \frac{1}{2} \frac{\mu}{r}.
\label{eq: dmu_dr}
\end{equation}
After integration, one can find that the viscous torque vanishes when $\mu \propto r^{-0.5}$. If the dynamic viscosity decreases with distance faster than $r^{-0.5}$, the viscous torque is positive (outward mass transport); otherwise, it is negative (inward mass transport).

\begin{figure}
    \centering
    \includegraphics[width=\columnwidth]{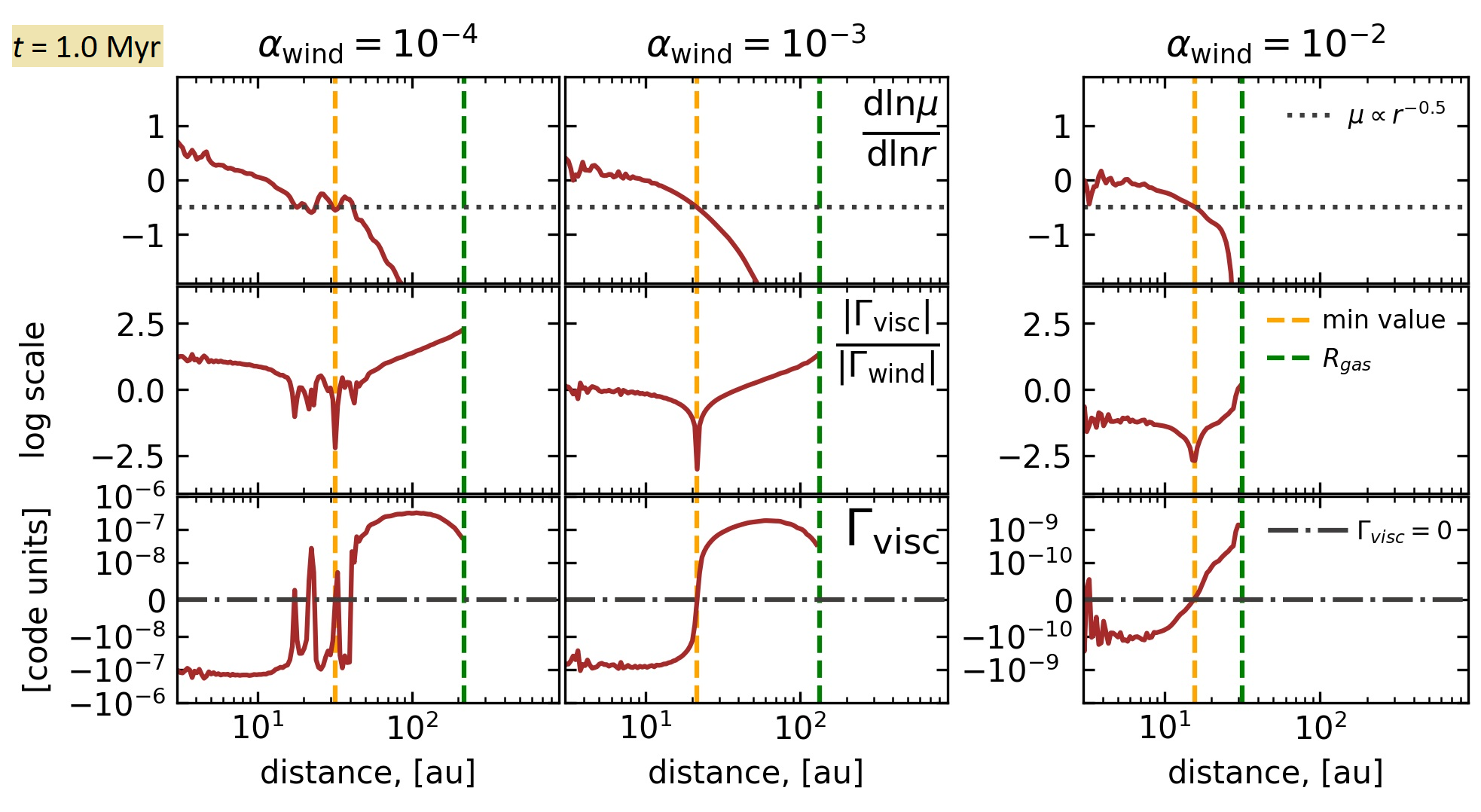}
    \caption{Radial profiles of azimuthally averaged quantities at time $t = 1~{\rm Myr}$: the slope of the dynamic viscosity (top panel), the decimal logarithm of the ratio of the absolute values of the viscous and magnetic torques (middle panel), the symmetric logarithm of the viscous torque in code units (bottom panel). The vertical dashed lines mark: the gas disc boundary (green) and the radial distance in the disc where the ratio $|\Gamma_{\rm visc}| / |\Gamma_{\rm wind}|$ is minimal (orange).}
    \label{fig: dlnMu_profiles}
\end{figure}

Fig.~\ref{fig: dlnMu_profiles} shows the radial profiles at time $t = 1~{\rm Myr}$ for all models with a magnetic disc wind. The top panel shows the profiles of the dynamic viscosity slope $d{\rm ln \mu} / d{\rm ln} r$, the middle panel shows the ratio of the absolute values of the azimuthally averaged viscous and magnetic torques (the temporal evolution of this quantity is represented by the color map in the middle panel of Fig.~\ref{fig: disk_torques_colormaps}), and the bottom panel shows the azimuthally averaged viscous torque in code units including the sign. For clarity, the profiles are truncated at the gas disc radius $R_{\rm gas}$ at this time (green vertical dashed line). It can be seen that at the radial distance where the most prominent decrease in the ratio $|\Gamma_{\rm visc}| / |\Gamma_{\rm wind}|$ is observed (orange vertical dashed line), the slope of the $\mu(r)$ profile is equal to $-0.5$, and the viscous torque changes sign from negative to positive. The latter is particularly evident in the model with $\alpha_{\rm wind} = 10^{-3}$.

\section{Impact of artificial viscosity}
\label{app: artificial_viscosity}

Artificial viscosity is introduced into the calculation to smooth out non-physical oscillations behind the shock front \citep{Neumann&Richtmyer1950}. However, the excessive influence of artificial viscosity can have a negative effect on the simulation results, since it is not a physical property of the medium.

\begin{figure}
    \centering
    \includegraphics[width=\columnwidth]{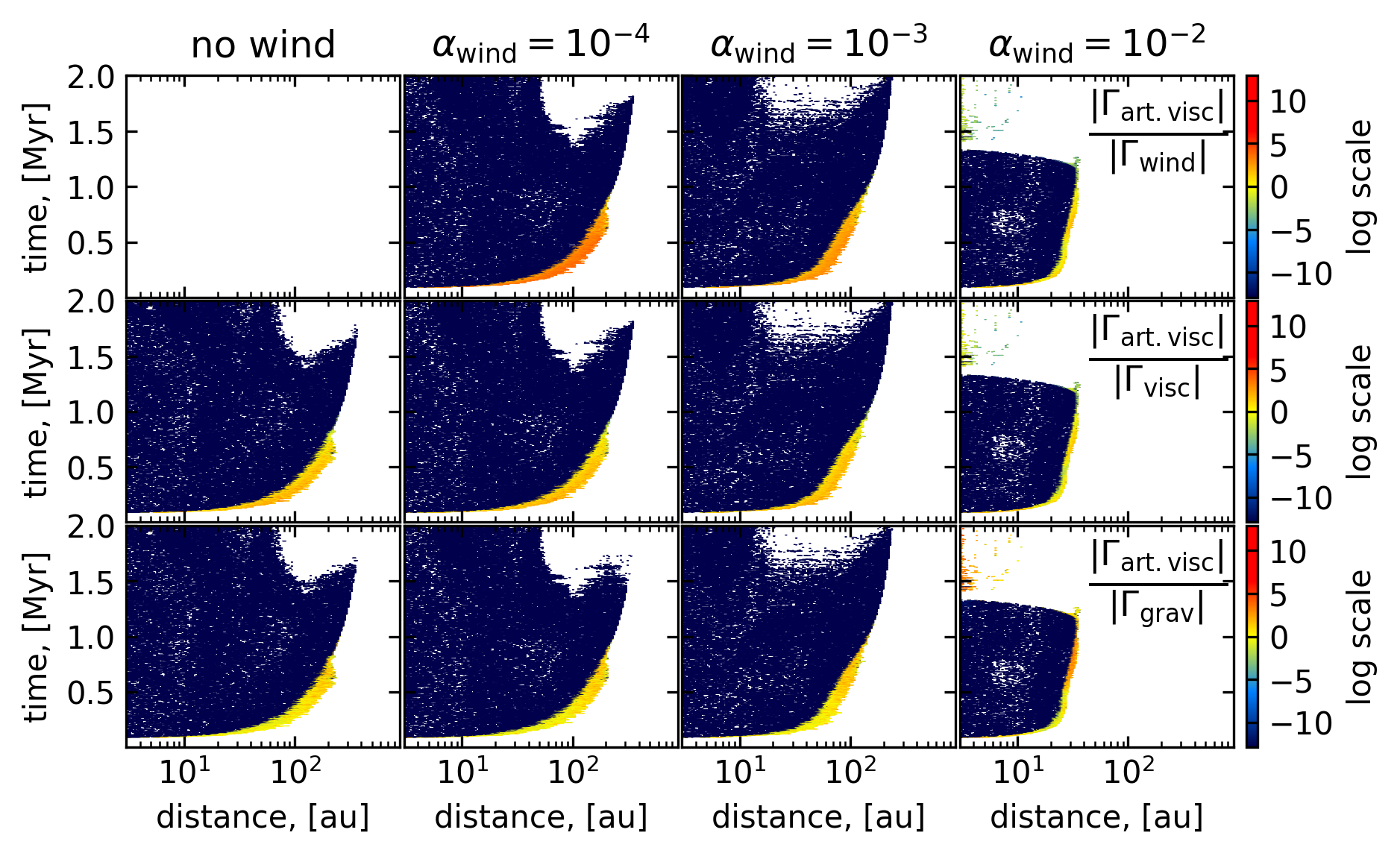}
    \caption{Temporal evolution of the ratio of the absolute value of the azimuthally averaged artificial viscosity torque to the absolute values of the azimuthally averaged magnetic, viscous, and gravitational torques (from top to bottom).}
    \label{fig: disk_torques_colormaps_with_aVisc}
\end{figure}

Fig.~\ref{fig: disk_torques_colormaps_with_aVisc} shows the time evolution of the ratio of the absolute value of the azimuthally averaged artificial viscosity torque to the other torques throughout the disc. As one can see in the figure, locally the artificial viscosity torque is comparable to or exceeds the other torques only at the disc-envelope interface during the first million years of evolution, where the collapsing envelope transitions into the protoplanetary disc and a shock wave arises. In the rest of the disc, the influence of artificial viscosity is negligible, and it does not affect the evolution.

\section{Magnitude of cumulative torque when summing local torques}
\label{app: torques_colormaps}

A conclusion about how the sign of the torque changes in the disc can be drawn by analysing the cumulative torques. Here, "cumulative" refers to the value of the azimuthally averaged torques summed in the radial direction. The sum is taken from the outer disc boundary towards the inner disc. The sign of the torque is taken into account during the summation.

\begin{figure}
    \centering
    \includegraphics[width=\columnwidth]{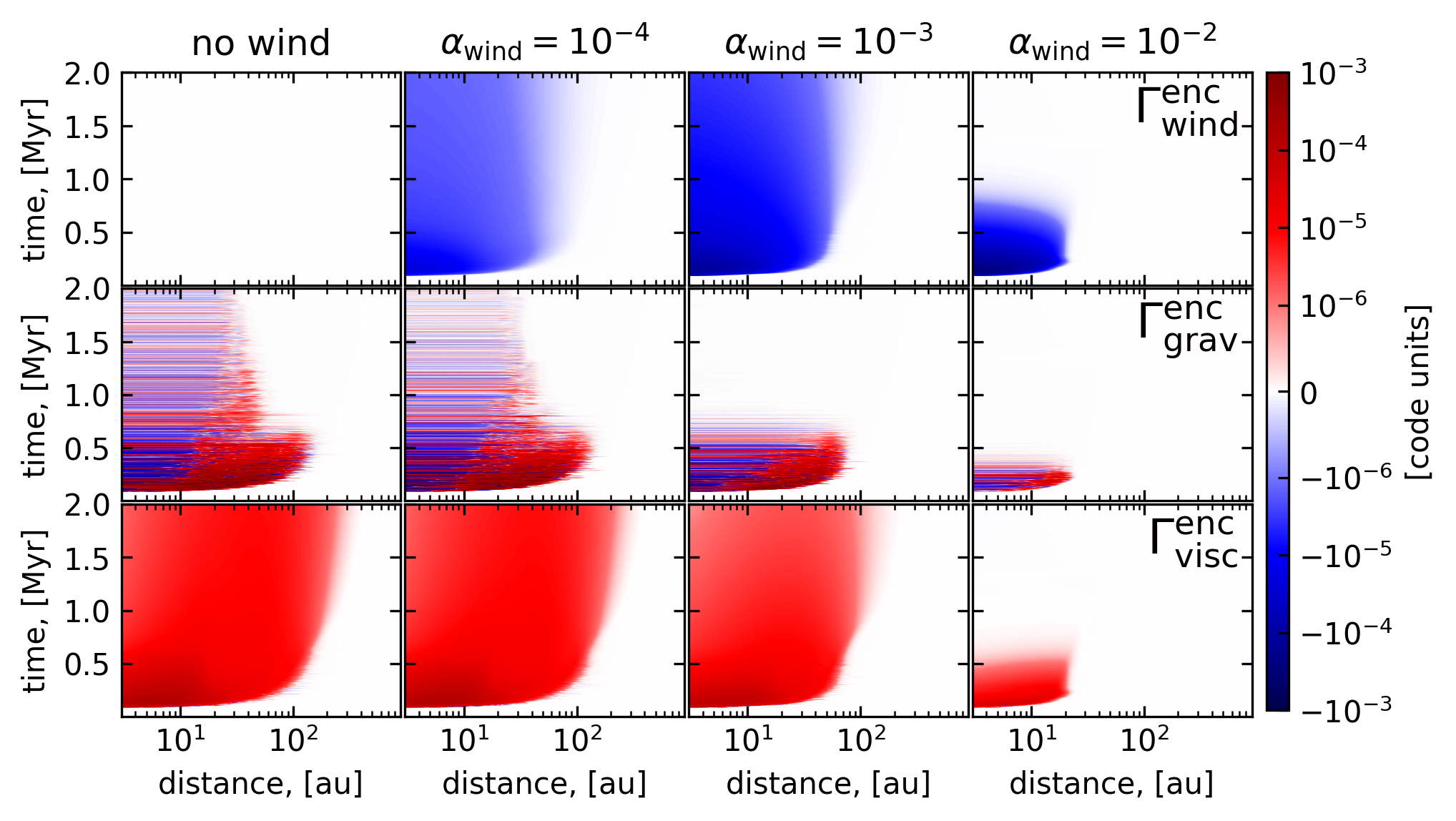}
    \caption{Temporal evolution of the cumulative torques. Top panel: magnetic torque, middle: gravitational, bottom: viscous. The sum is taken from the outer disc towards the inner region. The data are displayed in code units including the sign.}
    \label{fig: disk_torques_cummulative}
\end{figure}

Fig.~\ref{fig: disk_torques_cummulative} shows that the cumulative magnetic torque (top panel) is always negative, and as the inner boundary is approached, its absolute magnitude, as expected, increases, since all local magnetic torques are negative. The cumulative gravitational torque (middle panel) has predominantly positive values in the outer regions; however, as the inner boundary is approached, the negative local torques begin to prevail, and, as can be seen in most cases, the cumulative torque also becomes negative. A similar pattern is observed for the cumulative viscous torque. Although it always remains positive, it is evident that its magnitude, when moving towards the inner disc, first increases and then begins to decrease. This indicates that in the inner part of the disc the local viscous torques are negative.

\section{Evolution of mass fractions in models with different fragmentation velocity}
\label{app: mass_frac_v_frag}

\begin{figure}
    \centering
    \includegraphics[width=\columnwidth]{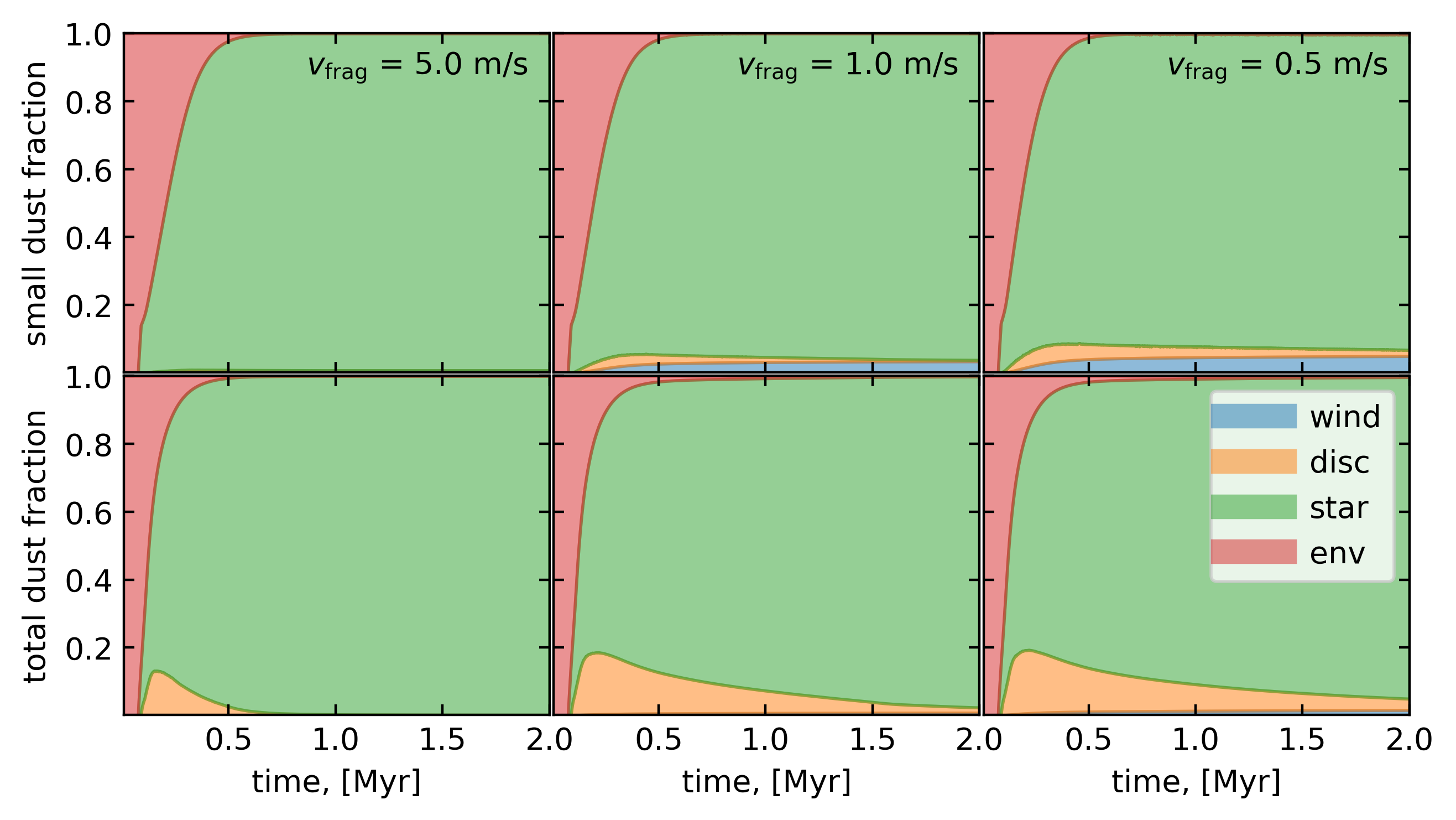}
    \caption{Evolution of the mass fraction of small (top panel) and total dust (sum of grown and small dust, bottom panel) contained in the envelope (red), on the star (green), in the disc (orange), and carried away by the wind (blue).}
    \label{fig: mass-gas-dust-fraction-vfrag}
\end{figure}

As described in Section~\ref{Subsection: dust_depletion}, lower values of fragmentation velocity $v_{\rm frag}$ result in a notable decrease in the maximum dust grain size, a process that is accompanied by the corresponding increase in the mass of small dust.  Fig.~\ref{fig: mass-gas-dust-fraction-vfrag} shows that in this case small dust particles are removed by the wind more efficiently. By $t$ = 2 Myr, in the model with $v_{\rm frag} = 0.5~{\rm m~s^{-1}}$, the mass fraction removed by the wind accounts for approximately 5.0\% of the small dust mass and 1.5\% of the total dust mass. In contrast, in the $v_{\rm frag} = 5~{\rm m~s^{-1}}$ model, 0.7\% of the small dust mass and 0.1\% of the total dust mass were carried away by the wind. At the same time, a smaller grain size in the $v_{\rm frag} = 0.5~{\rm m~s^{-1}}$ model results in a weaker inward radial drift, so the fraction of total dust retained in the disc remains higher as well: by $t$ = 2 Myr, the disc mass fraction is 3.37\% of the total dust mass, compared to only 0.02\% in the model with $v_{\rm frag} = 5~{\rm m~s^{-1}}$.

\section{Relations between wind parameters in different models}
\label{app: relations_between_wind_models}

When comparing the results presented in this work with those of the one-dimensional models of \citet{Suzuki2016} and \citet{Chambers2019}, it is useful to relate our parameter $\alpha_{\rm wind}$ \citep[and that of][]{Tabone2022} to the wind parameters used in those studies. \citet{Suzuki2016} use distinct parameters for the magnetic torque and the mass-loss rate due to the wind ($\overline{\alpha_{z \phi}}$ and $C_{\rm w}$, respectively), while in our work both processes are characterized by a single parameter $\alpha_{\rm wind}$. Comparing the mass-loss rate given in \citet{Suzuki2016}, $\dot{\Sigma}_{\rm w} = C_{\rm w} \rho_{\rm mid} c_{\rm s}$, with Equation (\ref{eq: sigma-dot-wind}), we obtain:
\begin{equation}
    \alpha_{\rm wind} = \frac{4 (\lambda - 1)}{3 \sqrt{2 \pi}} \left( \frac{H_{\rm g}}{r} \right)^{-2} C_{\rm w},
\label{eq: alpha_w_and_Cw}
\end{equation}
where it is taken into account that the midplane gas volume density is $\rho_{\rm mid} = \Sigma_{\rm g}/(\sqrt{2 \pi} H_{\rm g})$. Adopting the disc aspect ratio $H_{\rm g}/r \approx 0.05$ within inner $100$~au and $\lambda = 2$, an approximate relation between $\alpha_{\rm wind}$ and $C_{\rm w}$ can be written as:
\begin{equation}
    \alpha_{\rm wind} \approx 2 \times 10^2 \cdot C_{\rm w}.
\label{eq: alpha_w_and_Cw_simplified}
\end{equation}
Similarly, comparing their definition $T_{z \phi}^{\rm w} = \overline{\alpha_{z \phi}} \rho_{\rm mid} c_{\rm s}^2$ with our Equation (\ref{eq: alpha_wind}), we derive the relation between our $\alpha_{\rm wind}$ and their magnetic torque parameter, simplifying it for the case of $H_{\rm g}/r \approx 0.05$ and $\lambda=2$ as:
\begin{equation}
    \alpha_{\rm wind} = \frac{4}{3 \sqrt{2 \pi}} \left( \frac{H_{\rm g}}{r} \right)^{-1} \overline{\alpha_{z \phi}} \approx 10 \cdot \overline{\alpha_{z \phi}}.
\label{eq: alpha_w_and_alpha_phi_z}
\end{equation}

Using simplified relations (\ref{eq: alpha_w_and_Cw_simplified}) and (\ref{eq: alpha_w_and_alpha_phi_z}), as well as the relations between the parameters of \citet{Suzuki2016} and \citet{Chambers2019} written explicitly in the latter work, we can obtain the link between our parameter $\alpha_{\rm wind}$ and the parameters of \citet{Chambers2019}:
\begin{align}
    \alpha_{\rm wind} &\approx 10 \cdot \frac{f_{\rm w} v_0}{c_{{\rm s},0}}, \\
    \alpha_{\rm wind} &\approx 400 \cdot \frac{K f_{\rm w} v_0}{r_0 \Omega_0},
\label{eq: alpha_w_and_Chambers}
\end{align}
where the inward gas velocity $v_0$, sound speed $c_{{\rm s},0}$, and Keplerian angular velocity $\Omega_0$ are defined at the reference distance $r_0 = 1$~au for a temperature $T_0 = 150$~K; $f_{\rm w}$ is the fraction of the velocity $v_0$ caused by the magnetic disc wind; and $K$ is the parameter characterizing the mass loss due to the wind.

\end{document}